\documentstyle[aps,graphicx,epsfig,floats,tighten,preprint]{revtex}
\let\section=\subsection  \let\subsection=\subsubsection

\def\be{\begin{equation}}
\def\ee{\end{equation}}
\def\bea{\begin{eqnarray}}
\def\eea{\end{eqnarray}}
\def\Bphi{\mbox{\boldmath $\Phi$}}
\def\hphi{\mbox{\boldmath $\hat\Phi$}}
\def\vx{\mbox{\boldmath $x$}}
\def\Bxi{\mbox{\boldmath $\xi$}}
\begin{document}
\title{Features of phase ordering in (2+1)D-O(3) models }
\author{G. Holzwarth\thanks{%
e-mail: holzwarth@physik.uni-siegen.de} and J. Klomfass}
\address{Fachbereich Physik, Universit\"{a}t Siegen, 
D-57068 Siegen, Germany} 
\maketitle

\begin{abstract}\noindent
Numerical simulations of phase ordering under dissipative 
dynamics in a (2+1)-dimensional 3-vector model with $O$(3) symmetry are
reported. The energy functional includes terms which stabilize
the size of extended
topological defects. They emerge at the end of the coarsening process
as particle- or antiparticle-like structures floating in the globally
aligned vacuum. Approximate power-law growth of disoriented domains 
(with an exponent near 0.4) is found to be rather insensitive to the
size of the defects. An optional filter for conservation of
winding number allows to study phase ordering in high defect-density
environment which leads to large clusters of particles surrounded by
the aligned field.
\end{abstract} 

\vspace{3cm}
\leftline{PACS numbers: 05.45.-a,11.27.+d,12.39.Dc,73.40.Hm }  
\leftline{Keywords: Sigma models, Skyrmions, Bags, Disoriented domains,
Spin systems, }
\leftline{\hspace{ 19mm} Dissipative dynamics}
\newpage

\section{Introduction}
Effective field theories for order-parameter $N$-vector
fields can be powerful tools for the study of the dynamical behaviour of
complex microscopic many-body systems. Standard examples are
the local magnetization 3-vector field $\Bphi$ defined over (2+1)-dimensional
space-time for the description of pseudo-2-dimensional spin
systems in solid state physics~\cite{Burgess};
similarly, 4-vector fields defined over (3+1)-dimensional
space-time model low-energy QCD in chiral effective field theories
for the dynamics of pions and their interactions with gauge
fields~\cite{ChPT}. 
Extending these models to finite temperatures $T$ may 
allow to investigate features of ordering transitions in spin systems
or, for the hadronic case, in the cooling phase of the early universe
or immediately after a heavy-ion collision.

It is expected that during such a cooling process the system undergoes
a transition from a hot state in which the global symmetry of the
effective action is manifest into a cold state with spontaneously
broken symmetry, i.e. with global alignment of the ordering vector field
in a randomly chosen direction. The randomly oriented aligned domains
which characterize transient intermediate stages of this ordering
transition have recently found increased attention for the case of the
chiral meson field, because it was suggested that they might cause
anomalies in the multiplicities of emitted pions which could serve as a
signature for the phase transition itself~\cite{Anselm}. 
Similarly, the density of topological
defects in the field configurations has been related to production
rates of extended particle and antiparticle structures embedded in the
aligning field~\cite{Ellis}.

This latter aspect is based on the particularly interesting topological
properties of (2+1)D-$O$(3) and (3+1)D-$O$(4)  models. For appropriate
boundary conditions the field configurations fall into separate classes
which can be characterized by integer topological winding
number $B$. For $T=0$
the classical ground state configuration (the 'vacuum') and the 
structure of the particle-like excitations (the 'solitons' or
'baryons') are obtained through minimization of the corresponding
static energy functional $E[\Bphi]$. 
The vacuum consists of a globally aligned field with constant 
length $\Phi=f_0$ and $B=0$. In case of exact $O(N)$ symmetry different
global orientations are degenerate. Explicitly symmetry-breaking terms
or boundary conditions select a specific orientation as the true vacuum. 
Static topological defects are local minima of $E[\Bphi]$ in sectors
with $B\ne 0$. They represent ordered configurations with large gradients in
direction $\hphi$ and length $\Phi$ of the field vector. Large
deviations of $\Phi$ from the vacuum value $f_0$ are denoted as 'bags'. 
Stabilizing terms present in $E[\Bphi]$ determine soliton and bag
size, shape and spatial profile of isolated defects.

Such localized extended configurations have been
identified with particle-like structures, low-lying excitations with
smoothly varying magnetization in Quantum Hall Ferromagnets~\cite{QHE},
or baryons embedded in and interacting with the chiral mesonic
field~\cite{SkyWi}. In these cases the
corresponding local winding density has been interpreted as
local electric charge density, or baryon density, respectively.

Topologically nontrivial configurations can unwind at space-time points
where the length $\Phi$ of the field vector vanishes. So
configurations (with suitable boundary conditions) can be specified by a
definite winding number $B$ only if there is no point at which $\Phi=0$.
Representing local energy minima, static $B\ne 0$ configurations are
separated from global vacuum by barriers; they may be destabilized by
symmetry breakers which reduce these barriers. In time-dependent
(nonequilibrium) processes even for exact $O(N)$ symmetry 
$B$ might change through fluctuations or evolutions where $\Phi$ passes
through zero at some point. To study the dynamical behaviour of field
configurations at nonzero temperature for given average soliton density and
fluctuating $B$ therefore would require the inclusion of a chemical
potential and the construction of a large grandcanonical ensemble with 
numerous individual events in order to allow for a well-defined average
$B$-number. 
Alternatively, however, if different $B$ sectors are separated by
barriers, we can put a constraint on B. This is very naturally
implemented in lattice simulations if in a time evolution the
updates with occasional jumps in $B$ are rejected. Such a $B$-filter
then allows to investigate phase ordering in high defect-density
environment without the need for a grandcanonical ensemble.

The time evolution of classical fields on a lattice is considered as
the evolution of low-frequency modes subject to the noise of eliminated
high frequency fluctuations. In addition to stochastic forces this
provides a dissipative (first-order) time-derivative term which removes
energy from the field configurations and drives the system to the
thermal equilibrium at the temperature set by the noise term.
This temperature itself may change in time as a consequence of the
spatial evolution of the system, or as externally imposed quench.
Numerical simulations then may follow individual 'events' with
fixed winding number $B$ which start out from an initial random
configuration (which would represent a member of an ensemble 
of high temperature) with correlation length less than
the lattice constant. In such initial configurations the field vectors
vary randomly from one lattice point to the next, 
therefore these configurations are characterized by 
large local gradients, i.e. large (positive and negative) local winding
density, and large total energy. 

Ordering proceeds through growth of domains which comprise in their
interior increasing numbers of lattice vertices with aligned field
vectors. In the absence of any explicit symmetry breaking the relative
orientation of the aligned field for different domains is random. 
Therefore large field gradients then are confined to the boundaries of
these 'disoriented' domains. The soliton stabilizing terms in $E[\Bphi]$
then lead to a dynamical interplay between the growth of disoriented
domains and the formation of energetically favorable localized extended
structures. Finally, we expect a few well-developed solitons to remain
embedded in an otherwise fully aligned vacuum, each soliton with its own
(approximately integer) winding number $B_i$ which add up to the
conserved total integer winding number $B$. Residual interactions and
(at finite temperatures) remaining field fluctuations cause slow motion
of the localized structures such that on very long time scales some of
them may meet and combine or annihilate to form larger or smaller
$B_i+B_j$ structures.
 
Evidently, the whole evolution of one individual configuration is an
extremely complex process, and after all, only statistical statements
averaged over many events will be of interest. However, for a
sufficiently large lattice which finally still contains a large number
of individual solitons, lattice averages for individual events will
already provide good approximations to ensemble averages. In the present 
work we shall exclusively deal with the (2+1)D-$O$(3) model where
lattice sizes of the order of $10^2 \times 10^2$ are computationally
easy to handle and yet sufficiently large to observe essential features
of the ordering process.

Because the presence of topological textures implies the existence of
additional scales relevant for the time evolution it is expected that 
phase ordering in such systems violates dynamical scaling. This has 
been demonstrated in previous simulations for 2D-$O$(3)
non-linear~\cite{Zapo} and linear~\cite{Rut} sigma models.
In two spatial dimensions the two-derivative (sigma model) term
$\int\nabla\Bphi\nabla\Bphi d^2x$ is invariant with respect to the
spatial scale, therefore in these minimal models the size of the defects
is not fixed by the static energy functional. In fact, it was observed
in~\cite{Zapo,Rut} that the corresponding length scales change with time
and interfere with the correlation length scale that characterizes the
aligning process. 

It appears desirable to include in the energy functional 
additional terms which stabilize the defects at a fixed finite size.
Due to the scale invariance of the two-derivative term this requires at
least two more terms which balance each other in the stable static
configuration.
The effect of a four-derivative term
(which tends to increase the size of static structures) in the
nonlinear (hard-spin) version of the model can be
compensated by a (zero-derivative) Zeeman term which by itself tries to
shrink local inhomogenities. Physically motivated by external magnetic fields
coupled to the order field vector this term, however, explicitly breaks the
$O(3)$-symmetry and therefore prevents spontaneous alignment in random
directions. It is only in the easy plane where the formation of
disoriented domains can be observed as long as the field still has
components in that plane. Simulations of phase ordering in such models
have therefore mainly been concerned with the dependence of the 
resulting defect densities on the defect size~\cite{Ho}.

In the linear (soft spin) version of the model, however, where the
length $\Phi$ of the order-parameter field is not constrained, the
(zero-derivative) potential $V(\Phi)$ can serve instead to set the
length scale without breaking the $O(3)$ symmetry. 
With the familiar $(\Phi^2-f_0^2)^2$ ansatz for $V$
we arrive at a most simple model which allows for all features of
spontaneous symmetry breaking, combined with the possibility of
bag formation and the existence of localized structures with definite
size embedded in the aligning field.  
Inclusion of a four-derivative term is necessary to prevent the
collapse of these localized structures to zero size with subsequent 
unwinding.

It is the aim of this work to investigate phase ordering
in connection with the simultaneous formation of isolated
topologically nontrivial structures stabilized by the energy
functional. They appear as transient structures in every ordering
process, but here they finally persist as stable 'particles' and 
'antiparticles' with well-defined structure. We shall additionally 
consider the option that their net number is chosen as conserved
observable. 
In section 2 the effective lagrangian is specified 
which comprises the minimal number of terms necessary to establish
these features if we exclude all explicitly symmetry-breaking terms. 
In section 3 we briefly discuss the stable static solutions
of the corresponding energy functional which (in that model) only exist
if the relevant coupling constant is below a critical value.
Finally, in section 4 we perform numerical simulations of individual
'events' which follow the ordering process in real time through a
Langevin-type overdamped dynamics. For the exploratory purpose of this
work we only consider the sudden quench scenario where a random initial
configuration with correlation length less than the lattice constant
is exposed to a low-temperature effective potential and a
correspondingly small low-temperature stochastic force.
\newpage

\section{ The $2D$-$O(3)$ model with fourth-order stabilization }

We consider the $O(3)$-symmetric lagrangian density in $2+1$ dimensions
in terms of the dimensionless $3$-component field
$\Bphi = \Phi \hphi$ with $\hphi \cdot \hphi =1$,
\be\label{lag}
{\cal{L}} = F^2 \left(\frac{1}{2} \partial_\mu \Bphi \partial^\mu \Bphi
- \frac{\lambda_4}{4}\left( \Phi^2-f_0^2\right)^2
- c_4 \; \rho_\mu \rho^\mu \right).  
\ee
Apart from the usual sigma-model term this lagrangian
contains the standard potential $V(\Phi)$ for the modulus field $\Phi$
to monitor the spontaneous symmetry breaking,
and a four-derivative ('Skyrme') current-current coupling
$\rho_\mu\rho^\mu$ for the conserved 
topological current 
\be\label{top}
\rho^\mu = \frac{1}{8 \pi} \epsilon^{\mu \nu \rho} \hphi \cdot
( \partial_\nu \hphi \times \partial_\rho \hphi ) ,
\ee
which satisfies $\partial_\mu \rho^\mu = 0$.

If we write the independent strengths $\lambda_4$ of the
$\Phi^4$-coupling and $c_4$ of the Skyrme coupling in terms of one
common dimensionless parameter $\lambda$ and a length $\ell$ 
\be
\label{para}
\lambda_4=\lambda/\ell^2,~~~~~~~~~~~~~c_4=\lambda\ell^2
\ee
then $\ell$ may be
absorbed into the space-time coordinates. So, for $\lambda$ fixed, 
$\ell$ sets the size of localized static solutions,
and for continuous coordinates their total
energy is independent of $\ell$.
The overall energy scale is set by the parameter $F^2$.
Of course, we are free to insert additional powers of the modulus field
$\Phi$ into the Skyrme term, the above choice being
motivated to minimize interference with the $\Phi^4$ spontaneous 
symmetry-breaking mechanism. 

Having fixed the 
$\Phi$-dependence of the lagrangian as given in (\ref{lag}) and
(\ref{top}) we conveniently redefine the field and the parameters by 
\be
\label{rescale}
\tilde \Phi =\Phi f_0^{-1}, \qquad \tilde F^2=F^2 f_0^2, \qquad
\tilde \ell=\ell f_0^{-1}. 
\ee
This shows that for fixed $\ell$ as $f_0$ goes to zero ( e.g. with
increasing temperature ) 
the typical size $\tilde \ell$ of static defects grows like $1/f_0$.
This may be physically not unreasonable (cf. e.g. the discussion in
the 3-dimensional case in \cite{Hans97}). 
We omit the tildes in the following and absorb the $\ell$'s 
into the length scale of space-time. Then we finally have for the
static energy
\be
\label{energy}
E = F^2 \int \left(\frac{1}{2} \partial_i \Bphi \partial_i \Bphi
+ \frac{\lambda}{4}\left( \Phi^2-1 \right)^2
+\lambda \rho_0 \rho_0 \right) d^2x.  
\ee

For the following we will put the energy scale $F^2$ to unity. 
Note that the lagrangian (\ref{lag}) contains no symmetry-breaking term
and in this sense is the close analogue to the massless 
chiral 3D-$O$(4) model.

In the lattice implementation, we impose periodic boundary conditions 
for the field vectors which implies compactification of
coordinate space to a torus $S^1\times S^1$. A stronger condition would
be to require that $\Bphi$ is the same for all points on the lattice
boundary, which would imply compactification of coordinate space to
the two-sphere $S^2$. In both cases the winding density $\rho_0$
satisfies $\int \rho_0 d^2x =B$ with integer winding number B.

\section{Static soliton solutions}

Let us at first give a simple argument how for $\lambda$ less 
than a critical value  the formation of localized bags
will lead to field configurations with winding number $B$ which are
energetically more favorable than the standard
Belavin-Polyakov (BP) soliton solution~\cite{BP} of the $O(3)$ {\it
nonlinear} $\sigma$-model where $\Bphi$ is confined to the 2-sphere
$\Phi^2 \equiv 1$ everywhere.
For that purpose we consider idealized square-well bags, i.e.
configurations with fixed total winding number $B$ where 
$\Phi$ is close to zero inside an area $A$  and equals unity elsewhere,
with all nonvanishing angular gradients confined to the inside of that area.
For such configurations the winding density is $\rho=B/A$ and the first
term in (\ref{energy}) does not contribute, therefore
the bag energy (\ref{energy}) is minimal for $A=2B$ and is obtained as
\be
\label{ideal}
E_{bag}=\lambda B.
\ee
For $\lambda > 4\pi$ this exceeds the energy $E^{(BP)}_B=4\pi B$ of the 
standard BP solution , where the second term in (\ref{energy}) does
not contribute. In that case the contribution of the Skyrme term can be scaled
away by unlimited increase of the spatial scale. Therefore, for
$\lambda > 4 \pi$ idealized bags will not be stable but melt away into 
infinitly large BP-solitons. On the other hand, for
$\lambda< 4\pi$ we may expect well
defined stable bag structures with their spatial extent fixed by the
choice of $\ell$, for any chosen value of $B$.
\begin{figure}[t]
\begin{center}
\includegraphics[scale=0.45, angle=-90]{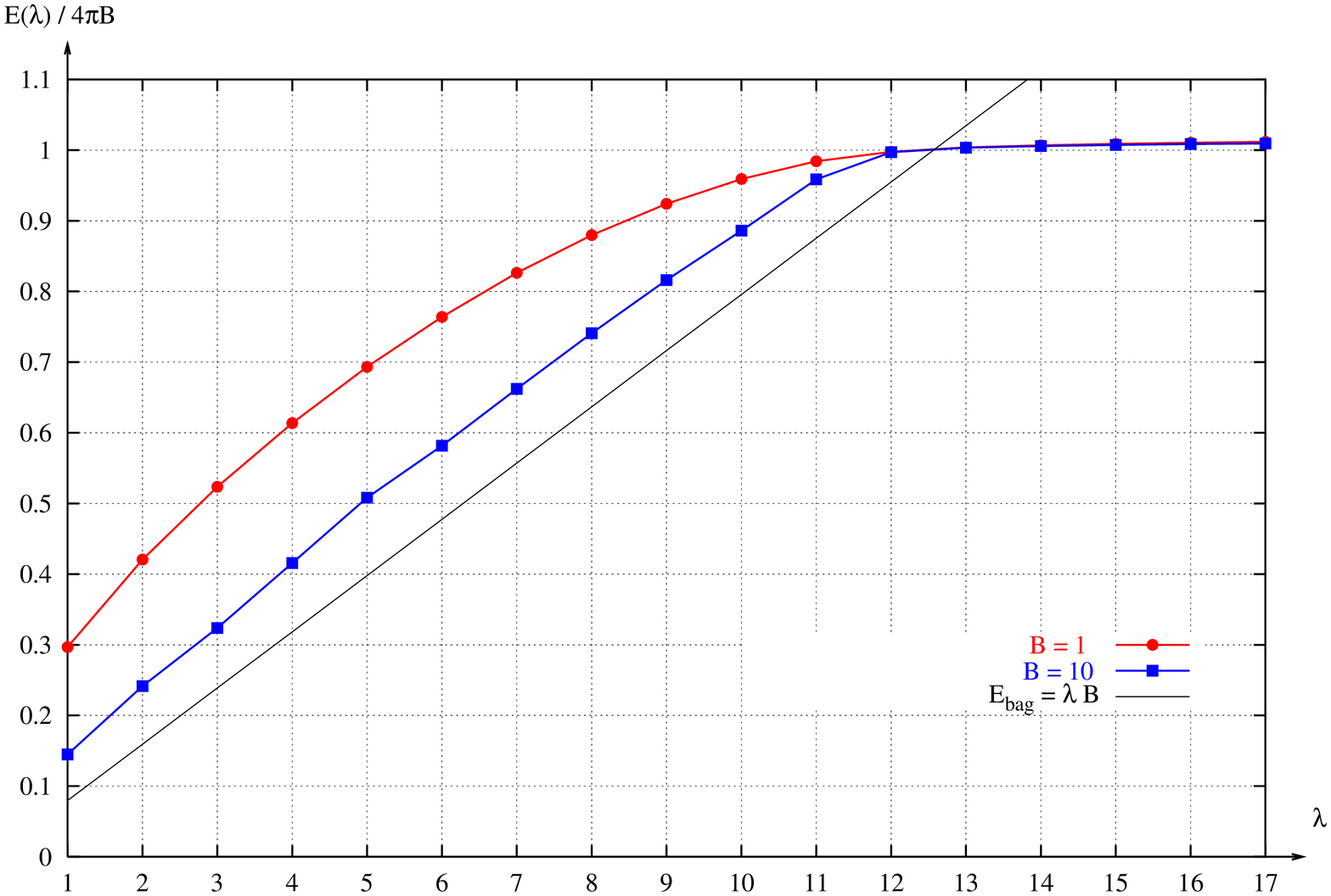}
\end{center}
\caption{Dependence of the total energy on the coupling strength
$\lambda$ for different values of $B$.} 
\label{Elambda}
\end{figure}

The above argument for idealized square-well bags,
involving only bulk energies, is independent of the shape of the
idealized bag. 
For real bags due to the surface energy given by the first term
in (\ref{energy}), the degeneracy of 
$E_{bag}$ with respect to the shape will be lifted.
With increasing values of $B$ these surface effects will be less and less
important.
This is shown in fig.\ref{Elambda}, where the energy $E( \lambda )$ is
plotted for $B=1$ and $B=10$.
As expected, the corresponding energy curves $E(\lambda)$ lie slightly
higher than the linear result (\ref{ideal}) derived above for idealized
square-well bags, and approach that result with increasing values of $B$.

In numerical simulations on a discrete $N\times N$ lattice the lattice
constant $a$ defines an additional
 scale so we can expect independence of the
energy from the scale $\ell$ only 
as long as $N \gg \ell/a = (c_4/\lambda_4)^{1/4} \gg 1$.
Then the energy $E(\lambda_4,c_4)$ resulting from the static part of
(\ref{lag}) will scale as $E(\lambda)$ in (\ref{energy}) of the single argument
$\lambda = \sqrt{\lambda_4 c_4}$, only.

\begin{figure}[h]
\centering
\includegraphics[scale=0.45, angle=-90]{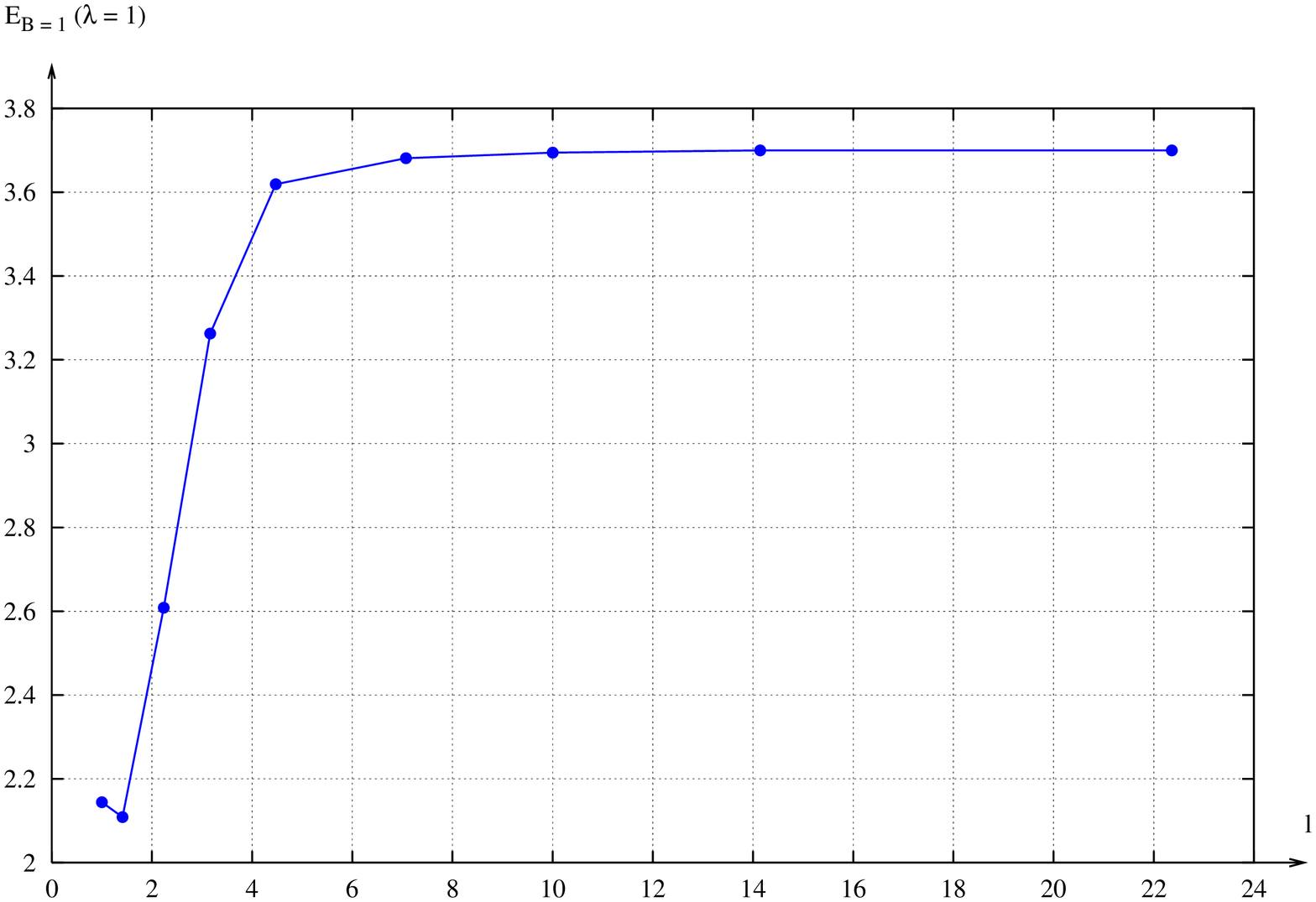}
\caption{The $\ell$-dependence of the energy $E_{B=1}$ 
for fixed coupling constant $\lambda = 1$.}
\label{scale}
\end{figure}

As $\ell/a$ approaches 1 from above scaling violations set in.
This is illustrated in fig.\ref{scale} where $E_{B=1}(\lambda=1)$ obtained on
a square lattice with $a=1$ is plotted for different values of $\ell$.  
Fig.\ref{scale} shows that scaling holds with good accuracy for $\ell> 4$.

In the angular representation of the field 3-vector $\Bphi = \Phi \hphi$
the length $\Phi$ is the 'bag' field, and it is convenient to
parametrize the angular part $\hphi$ in terms of the 
profile function $\Theta(x,y)$ and the azimuthal angle $\phi(x,y)$
with respect to some arbitrarily chosen cartesian basis
\be
\left(\Phi_1,\Phi_2,\Phi_3\right)=
\Phi\left(\cos\phi \sin\Theta,\sin\phi \sin\Theta,\cos\Theta\right).
\ee

\begin{figure}[h]
\begin{center}
\includegraphics[width=7.1cm, height=13.0cm, angle=-90]{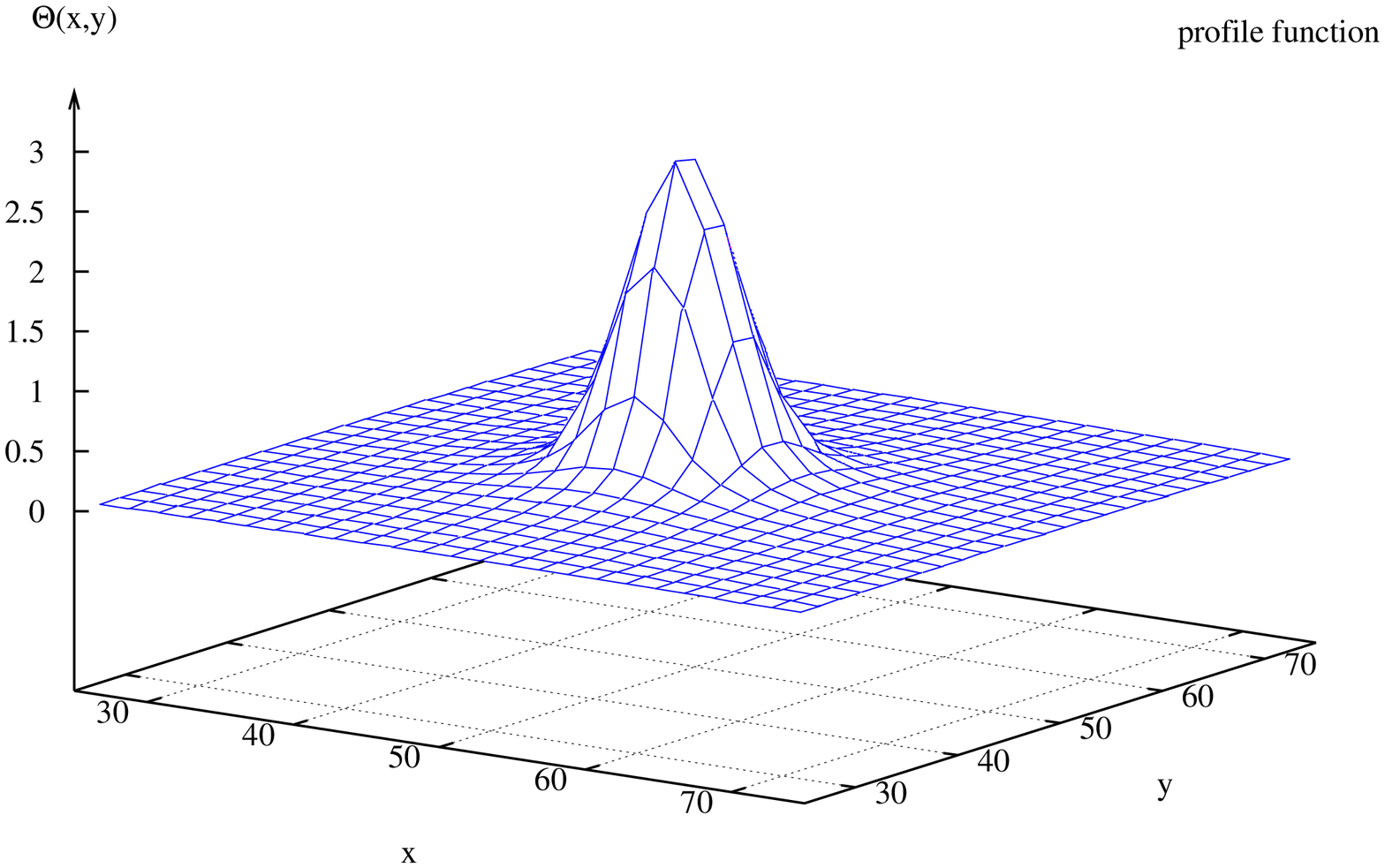}
\includegraphics[width=7.1cm, height=13.0cm, angle=-90]{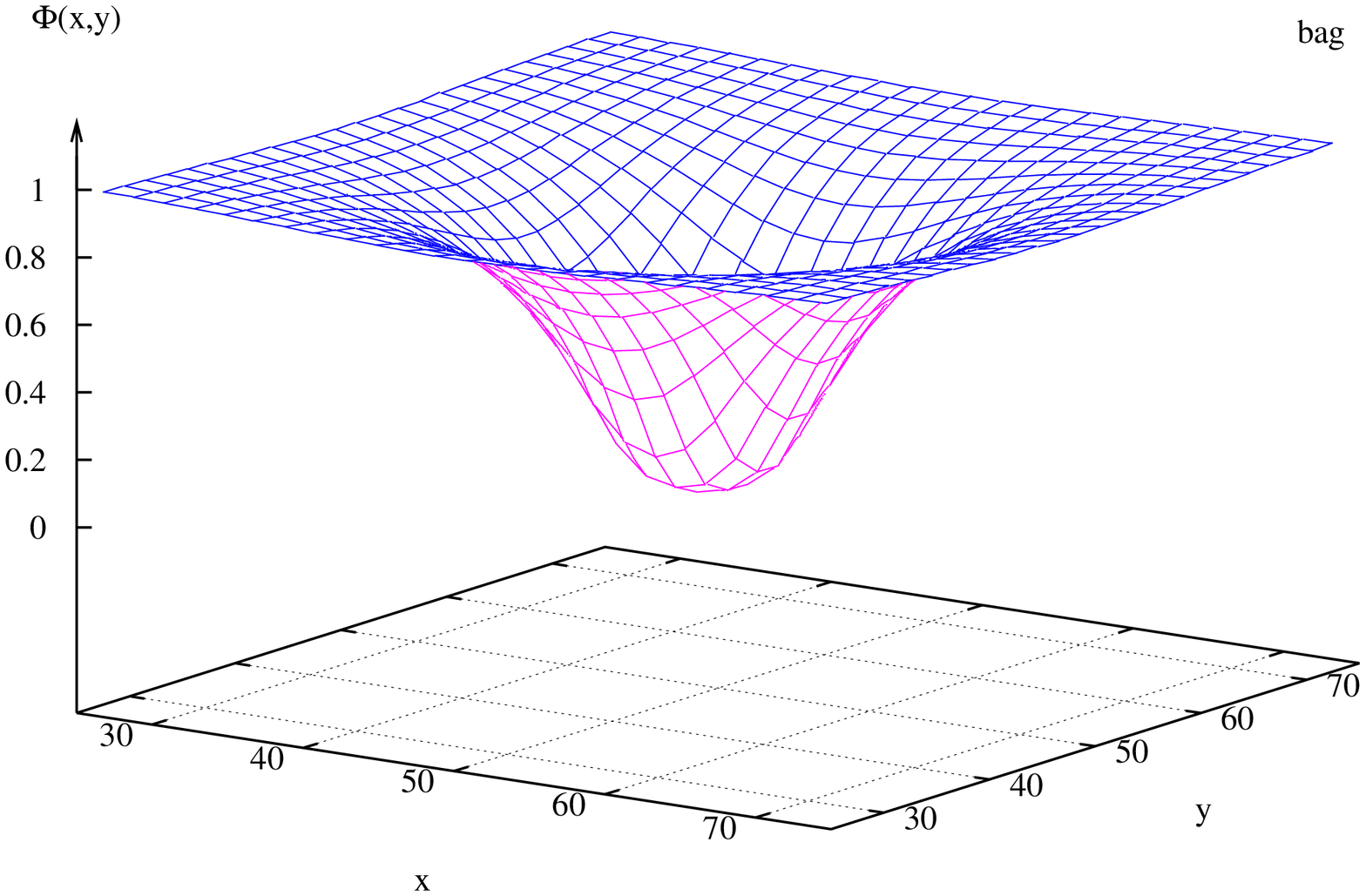}
\includegraphics[width=7.1cm, height=13.0cm, angle=-90]{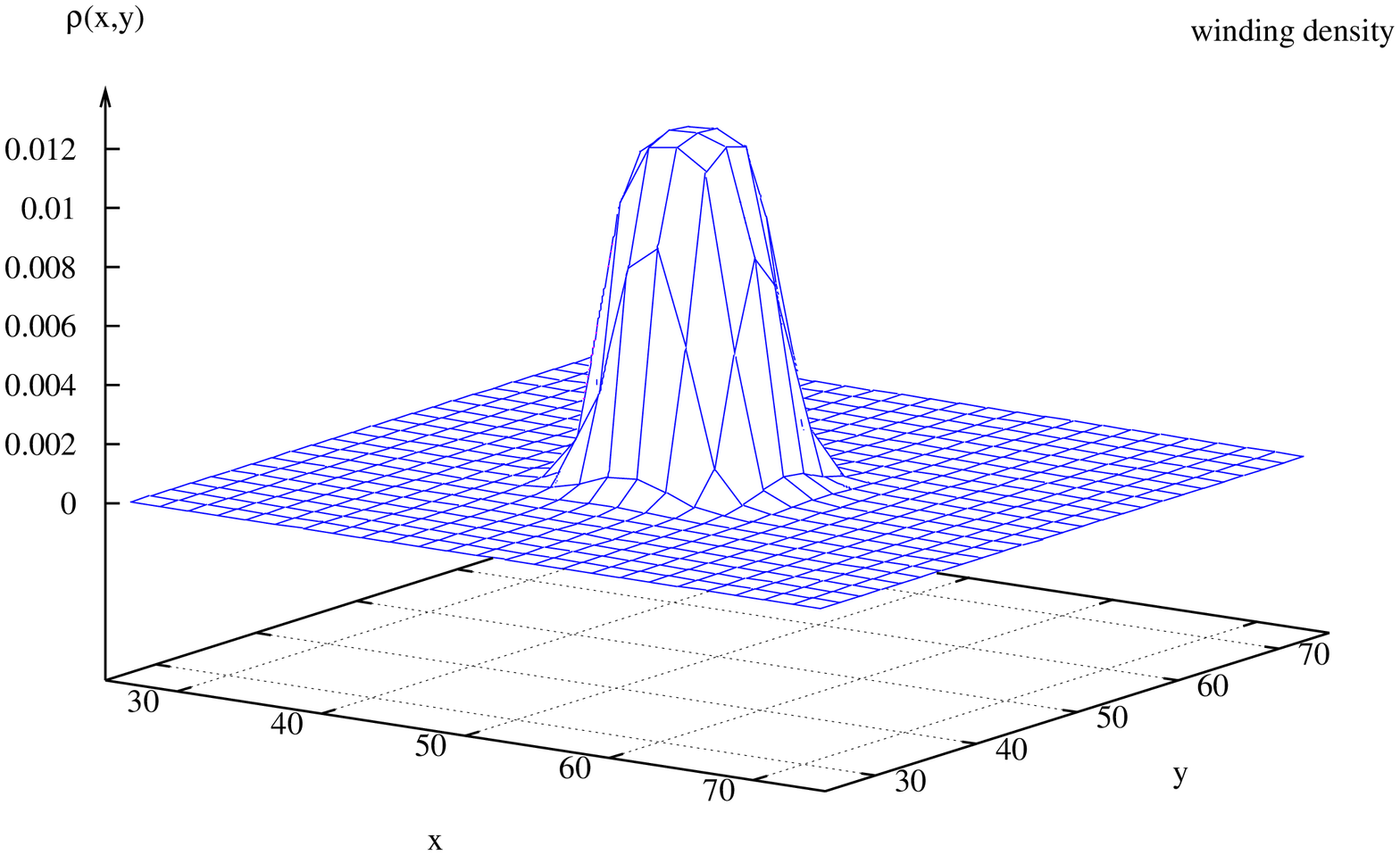}
\end{center}
\caption{Profile function $\Theta(x,y)$,  bag $\Phi(x,y)$, and the
winding density $\rho(x,y)$ of the $B=1$-configuration. ($\lambda=1$)}
\label{fields}
\end{figure}

For $\ell=10$, i.e. well within the scaling region,
the bag $\Phi(x,y)$, the profile function $\Theta(x,y)$, and the
winding density $\rho(x,y)$ of the resulting $B$=1$-$configuration are 
plotted in figs.\ref{fields}. 
One may recognize how the winding density  is concentrated within
the well-developed bag. The 'profile'-function $\Theta$
drops from the value of $\pi$ in the center to zero outside the bag
and the angular field $\phi(x,y)$
coincides with the BP-hedgehog form $\phi=\arctan(y/x)$. So, although
the angular configuration resembles closely the BP-soliton the energy
$E_{B=1}$ is (for $\lambda$=1) only 3.70 as compared to $E^{(BP)}_{B=1}=4\pi$.
\clearpage

The effects of the finite lattice constant on the field configurations
can be studied
as $\ell$ approaches 1 from above: the bagfield $\Phi$ 
develops a sharp dip by taking on
a value very close to $\Phi=0$ only at one single lattice point while being a
smooth extended function otherwise. The
corresponding density $\rho$ for a configuration with $B=1$ 
then assumes the values of $\rho=0.25$
within each of the four adjacent lattice cells with $\rho\equiv 0$ everywhere
else. Then, while the bagfield still is able to scale with $\ell$ as 
$\Phi_\ell(\vx)=\Phi(\vx/\ell)$, the density $\rho^0$ can no longer scale as  
$\rho_\ell(\vx)=\rho(\vx/\ell)/\ell^2$. This then causes the
scaling violations in the energy $E_{B=1}$ as shown in fig.\ref{scale}.
If $\ell$ is chosen still smaller, like $\ell = 0.1$,
the density takes the value of 0.5 on two adjacent lattice cells
while the bag zero disappears somewhere between the lattice points, such
that the field $\Phi \approx 1$ on all lattice vertices. In that case the
lattice simulation produces a configuration which even looks as if the
constraint $\Bphi^2 \equiv 1$ had been imposed.
In that case large bags carrying multiple charges break up into
individual $B=1$ structures. 
In order to avoid such effects of the finite lattice constant
we consider in the following scales $\ell$ which are
sufficiently large to be safely in the
scaling region but still small enough for the resulting configurations
to be well contained in a reasonably sized lattice (like $N=100-150$).

\begin{figure}[t]
\begin{center}
\includegraphics[scale=0.50,angle=-90]{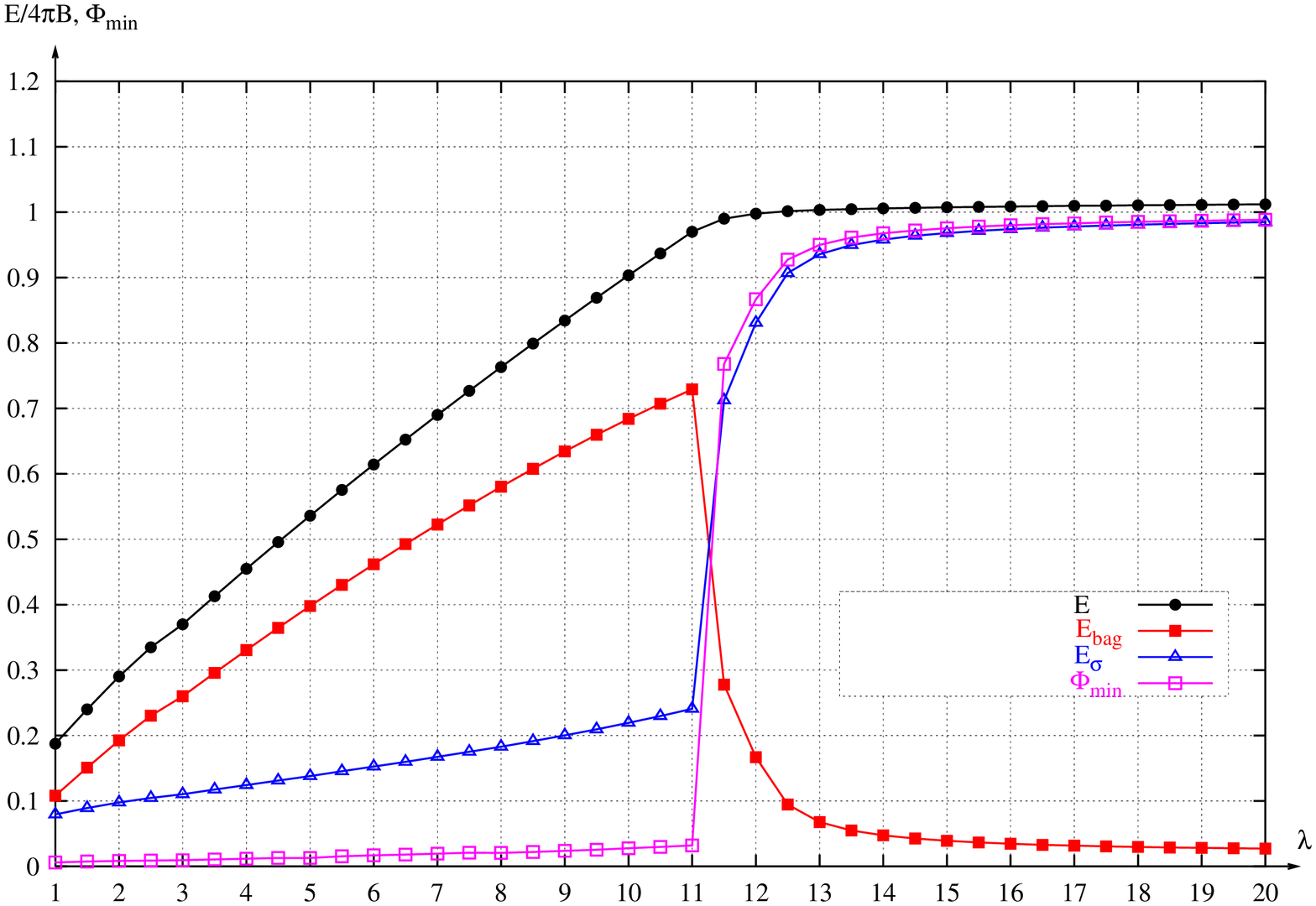}
\end{center}
\caption{
Total energy $E = E_{\sigma} + E_{bag}$, 
surface energy $E_{\sigma} = \frac{1}{2} \int \partial_i \Bphi
\partial_i \Bphi \ d^2x$, 
bag energy     $E_{bag} = \lambda \int \left( \frac{1}{4}\left(
\Phi^2-1 \right)^2 
+ \rho_0 \rho_0 \right) d^2x$,
and the minimum of the 'bag'-field $\Phi_{\rm min}$,  as  functions of
the coupling constant 
$\lambda$, for $B=4$.}
\label{Econtrib} 
\end{figure}

\begin{figure}[h]
\begin{center}
\includegraphics[scale=0.44, angle=-90]{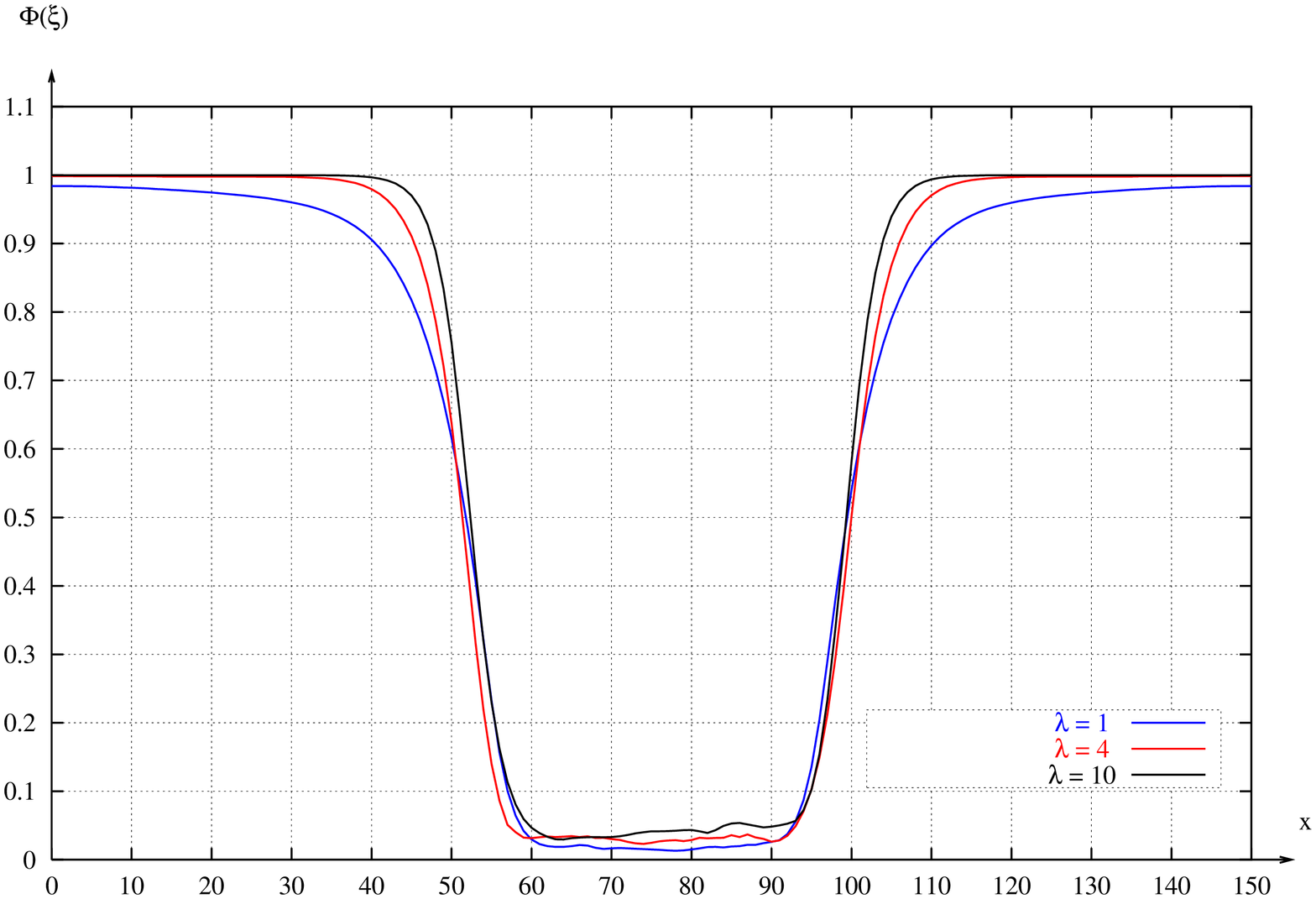}
\includegraphics[scale=0.44, angle=-90]{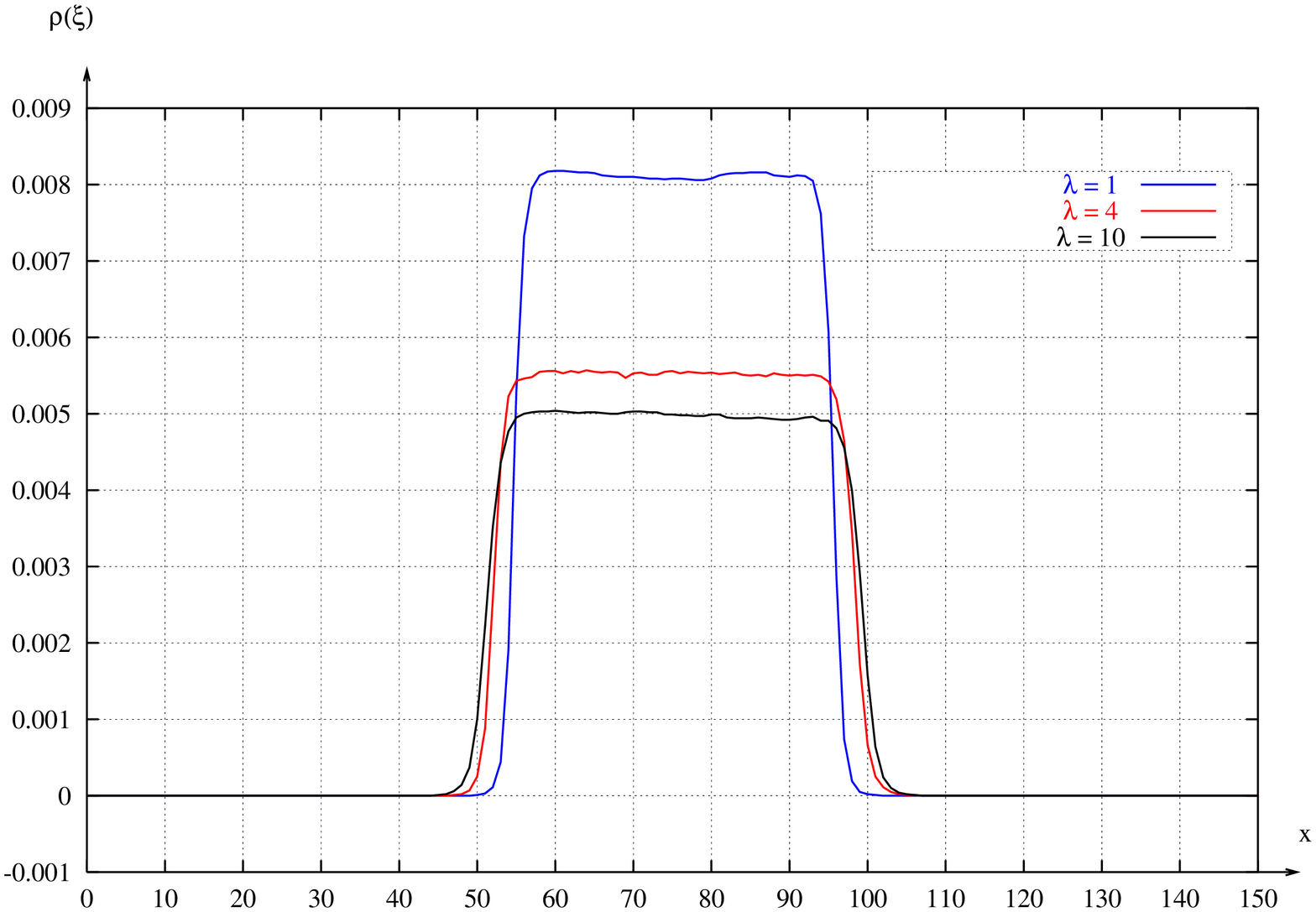}
\end{center}
\caption{Density and  bag profiles for $B=10$ and 
three different values of $\lambda=1,4,10$. ($\ell = 10$) }
\label{bags}
\end{figure}

We finally proceed to the dependence of 
stable bag configurations and their energies on the coupling strength
$\lambda$. 
For $B=4$ different contributions to the total energy and the 
depth of the bag profile are shown in fig.\ref{Econtrib}, and in
figs.\ref{bags} densities and bag profiles for $B=10$ and
$\lambda=1,4,10$ are compared.
Referring to fig.\ref{Econtrib},
one may still recognize nonvanishing
bags for $\lambda > 4 \pi$ as a consequence of the limitation of the
lattice size: 
The size of the BP-soliton is restricted by the borders of the
lattice, so that the contribution of the Skyrme term cannot really
be scaled away for $\lambda > 4 \pi$. 
According to Hobart-Derrick's theorem the second and the
third term in (\ref{energy}) must contribute
the same amount to the total energy,
resulting in very flat but still nonvanishing bags for $\lambda>4 \pi$.
Figs.\ref{bags} show that for $B=10$ the bags already resemble the idealized
bags discussed above quite closely, with a flat interior in which the
similarly flat density is localized. 
Their radius is almost independent of the
coupling constant $\lambda$ and fixed by the choice of $\ell$.
The surface thickness depends on $\lambda$ but not in a dramatic
way. Mainly the outermost tails of the bag profile are 
sensitive to $\lambda$. However, increasing surface thickness squeezes
the winding density towards the center of the bag such that the central
value of the local density can be quite sensitive to $\lambda$,
especially for small values of $\lambda$.   
\clearpage

\section{ Relaxation after sudden quench}

Overdamped relaxation of initially random configurations leads to
the formation of domains in which the field vectors are aligned in
spontaneously chosen random directions. Size and orientation of these
(dis)oriented domains change with progressing time, some of
them growing on cost of others, such that altogether long-range order is 
increasing. Boundaries and edges of such domains are characterized by
large angular field gradients. Energetically, large angular gradients 
favor formation of 
bags where the length of the field vectors deviates strongly from the 
vacuum value $f_0$. Therefore the ordering process is accompanied by
spontaneous formation of bags with winding density accumulated inside
the bags. With progressing time these bags assume the spatial extent
and profile dictated by the energy functional, while the areas of
aligned field in which they are embedded finally grow and coalesce 
into a uniformly oriented vacuum.

In order to follow this ordering process as it proceeds in real time
we consider the equations of motion as obtained from (\ref{lag}),
suppressing, however, all second-order time derivatives in comparison to
a first-order time-derivative dissipative term:
\be
\label{eom}
\frac{1}{\tau}\dot{\Bphi}= \Delta\Bphi
-\frac{\lambda}{\ell^2} \left( \Phi^2-1 \right) \Bphi
-\lambda \ell^2\; \partial(\rho_0^2) /\partial \Bphi + 
\frac{1}{\ell^2}\Bxi.
\ee
As long as we disregard second-order time-derivatives the damping constant
$1/\tau$ which multiplies the $\dot{\Bphi}$ term can be 
chosen as unity, i.e. we identify the time unit with the relaxation
time $\tau$.  

In (\ref{eom}) we also have added a fluctuational field $\Bxi(\vx,t)$
to represent gaussian white noise; in principle its 
presence and strength is dictated by the dissipation-fluctuation
theorem. Independent from this stochastic dissipation other
damping mechanisms could be present, like the rapid cooling due to
the Bjorken expansion of a hot hadronic fireball~\cite{Bjorken}.
So, depending on the specific physical situation the damping rate and
other parameters in (\ref{eom}) may be subject to an appropriate
time (or temperature) dependence. 
Here, however, for definiteness, we will keep them fixed during each
individual evolution. This corresponds to a sudden quench where the
initially (at $t=0$) hot configuration is exposed for $t > 0$ to the
low-temperature ($T=0$) effective action. Consequently, we also
generally will omit the noise term $\Bxi$. Average results are not 
sensitive to it, anyway; only the accidental features of late-time
configurations reached in individual evolution events are affected
by the noise term.

Initial configurations are chosen such that at each lattice vertex
$(i,j)$ $(i,j=0,..,N)$ 
the field vectors $\Bphi$ point in some random direction 
i.e. at each point of the lattice the angle $\phi(i,j)$ is selected
randomly from the intervall $[0,2\pi]$, the angle $\theta(i,j)$ from
the intervall $[0,\pi]$. Through this choice the finite lattice
constant acquires physical meaning as providing a measure for the 
magnitude of the initial correlation length. The moduli $\Phi(i,j)$ of
the field vectors $\Bphi$ are chosen as absolute values of 
a Gaussian deviate around the symmetry center $\Phi=0$. 
Again, late-time average features of the resulting configurations do
not depend significantly on the mean square deviation of this initial
gaussian distribution. In fact,
similar results are obtained even if the initial configuration is
constrained to the 2-sphere $\Phi^2=f_0^2$. The reason for this is that
the system reacts to the initially large local angular gradients by
reducing the length of the field vectors almost everywhere to values
which are small as compared to $f_0$. This happens early during the
first few time steps, accompanied by some next-neighbor alignment.
Therefore the initial length distribution is almost instantly forgotten.

At the borders of the lattice periodic boundary conditions
are enforced. If we divide each elementary lattice cell with lower left
corner $(i,j)$ into two triangles (e.g. by the same diagonal in all
cells), then 
the map $\hphi(i,j)$ maps each triangle onto a spherical triangle on the 
sphere $\Phi^2 =1$ cut out of the surface of this sphere by the 
(shortest) geodesics which connect the image points of the corners of
each triangle. The local winding density $\rho(i,j)$ then is defined 
as the sum of the (oriented) areas (divided by $4\pi$) of the two 
spherical triangles which
form the images of the lattice cell with lower left corner $(i,j)$. 
Because the area of each spherical triangle is less than $2\pi$ each
square lattice cell can contain at most one unit of total winding number. 
The periodic boundary conditions guarantee that the total
winding number
\be
B=\sum_{i,j=0}^{N-1} \rho(i,j)
\ee
summed over the whole lattice is integer, so that initial configurations can
be selected with some desired integer value of $B$. We also define the
"number of defects"
\be
\label{D}
D=\sum_{i,j=0}^{N-1} |\rho(i,j)|
\ee
by summing up the absolute values of the local winding densities.
Of course, for random or slowly varying smooth configurations
$D$ generally is not an integer, but if a configuration describes
a distribution of localized defects (and antidefects) which are
sufficiently well separated 
from each other, then $D$ is close to an integer and counts the number
of these defects (plus antidefects). In that case we can define the
numbers $N_+,N_-$ 
of "particles" and "antiparticles" through
\be
B=N_+ - N_-~~~~~~~~~~~D=N_++N_-~~.
\ee
We shall, however, in the following (sloppily) call $D$ the "particle number"
(even if it is not integer). (Alternatively, $D$ could be defined as
the sum of the absolute values of the areas of all spherical triangles
considered above. For a random configurations this would result in an
average value of $\langle D\rangle=N^2/4$ (Kibble limit~\cite{Kibble}).
Our definition (\ref{D}) for random configurations leads to 
 $\langle D\rangle=0.73 N^2/4$ which implies a slightly different
definition of the initial correlation length.)

The total winding number $B$ always is integer and occasionally will
undergo discrete jumps in the update sweeps. For well-developed
localized structures this corresponds to unwinding defects or
antidefects independently, such that $B$ decreases or increases by one
or more units. This $B$-violating propagation is
characteristic for the trivial topology of the linear $O(3)$-model.
However, with the evaluation of $B$ for each instantaneous
configuration we may in the lattice simulation implement an (optional)
$B$ filter which in each time step rejects configurations that violate $B$
conservation. This eliminates all independent unwinding processes. Only
simultaneous annihilation of defect and antidefect in the same time
step remains possible, and, as the update proceeds locally at each
lattice vertex it can happen only if defect and antidefect overlap.
This $B$-conserving evolution is characteristic for the nontrivial
topology of the nonlinear $O(3)$-model. 
Of course, we expect that severe differences between both types
of evolutions appear only if $D$ is comparable to $B$.

In order to produce configurations with well-developed stable bags
we choose for the following quench simulations 
values for the coupling constant between $0.5<\lambda<5$. 
This is well below the critical value of $\lambda = 4 \pi $.

For most of the results presented below the scale $\ell$ is chosen
as $\ell=\sqrt{10}$; this is small enough to allow for the formation of 
numerous bags on a reasonably sized lattice (like $150\times 150$),
but is still close to the onset of the scaling region (cf.
fig.\ref{scale}) to suppress effects of the finite lattice constant. 

\begin{figure}[t]
\begin{center}
\includegraphics[scale=0.45, angle=-90]{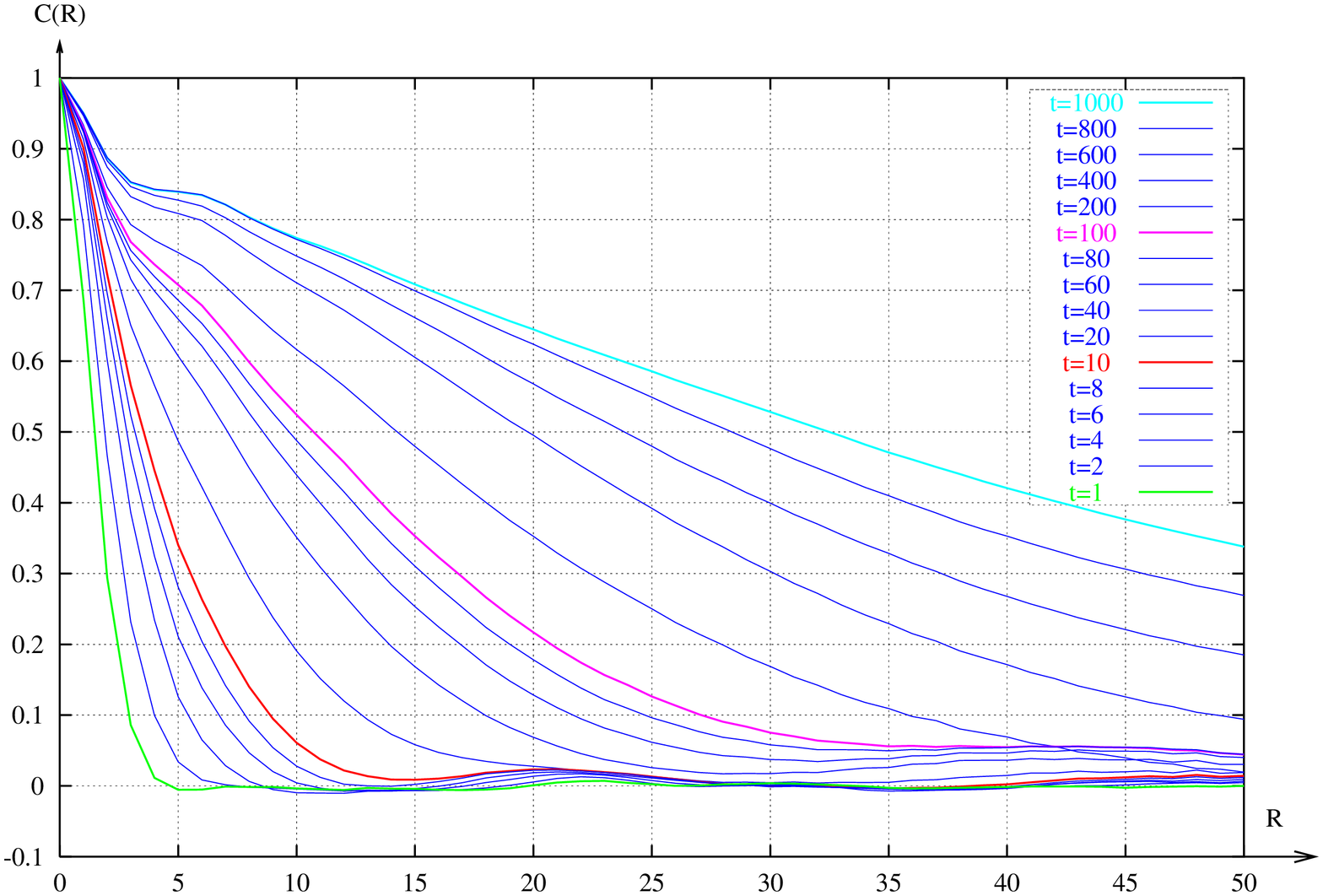}
\end{center}
\caption{Time evolution of the correlation function  for $B = 0$ and
$\lambda=5$. The time is given in units of the relaxation time $\tau$.} 
\label{cor0}
\end{figure}
To obtain a more quantitative measure for the size of
ordered domains we consider the correlation function
\be
\label{corr}
C(R)=\sum_{i,j=0}^{N}\sum_{k,l=0}^{N} \hphi(i,j)\cdot \hphi(k,l)
/\sum_{i,j=0}^{N}\sum_{k,l=0}^{N} 1
\ee 
where the $k,l$ sum is restricted such that 
the distance $r=\sqrt{(k-i)^2+(l-j)^2}$ between lattice vertices
$(i,j)$ and $(k,l)$ lies inside bins of unit size around
fixed positive integers $R$. The typical shapes of these correlation functions
are shown in fig.\ref{cor0}  for an evolution with
$B$-conservation on a 150$\times$150 mesh, for increasing time.
For the initial configuration $C(R)$ vanishes for all $R \ge 1$
which reflects our choice of the initial correlation length. 
In the very early part of the relaxation process ($t < 1$)
these correlation functions approach zero within less than 5 lattice
units and stay close to zero for larger distances. For $t > 10$
they drop to small values above 10 lattice units, but stay
positive with small oscillations ( with wavelengths of more than 20
lattice units, increasing with $t$). For $t>200$
these oscillations no longer fit into the 150$\times$150 lattice,
the correlation functions show a monotonous decrease (for $R<N/2$).
Their minima (near $R = N/2$) increase towards unity, which
indicates that finally the extension of the ordered domains reaches
the size of the lattice.

Comparing with typical field configurations during the relaxation,
the half-maximum distance, i.e.
the distance $R$ where $C(R)$ drops below $0.5$ apparently provides an
appropriate measure for the 'radius' $R_D$ of ordered domains.
Of course, this is a
rather arbitrary and not very precise convention, but it captures the 
essentials of the ordering process in view of the fact that the
boundaries of the ordered domains are not sharply defined.
A typical feature of these correlation functions is the appearance
of a shoulder for small distances ($R<5$) for late times. This reflects
the formation of the ordered textures, i.e. spatially extended
angular twists which locally prevent alignment of the field vectors
over distances of 
the order of $\ell$. Naturally, this effect gets especially prominent for 
evolutions which proceed in configurations constrained to a large
total winding number $B$ (see below). 

\begin{figure}[t]
\begin{center}
\includegraphics[scale=0.45, angle=-90]{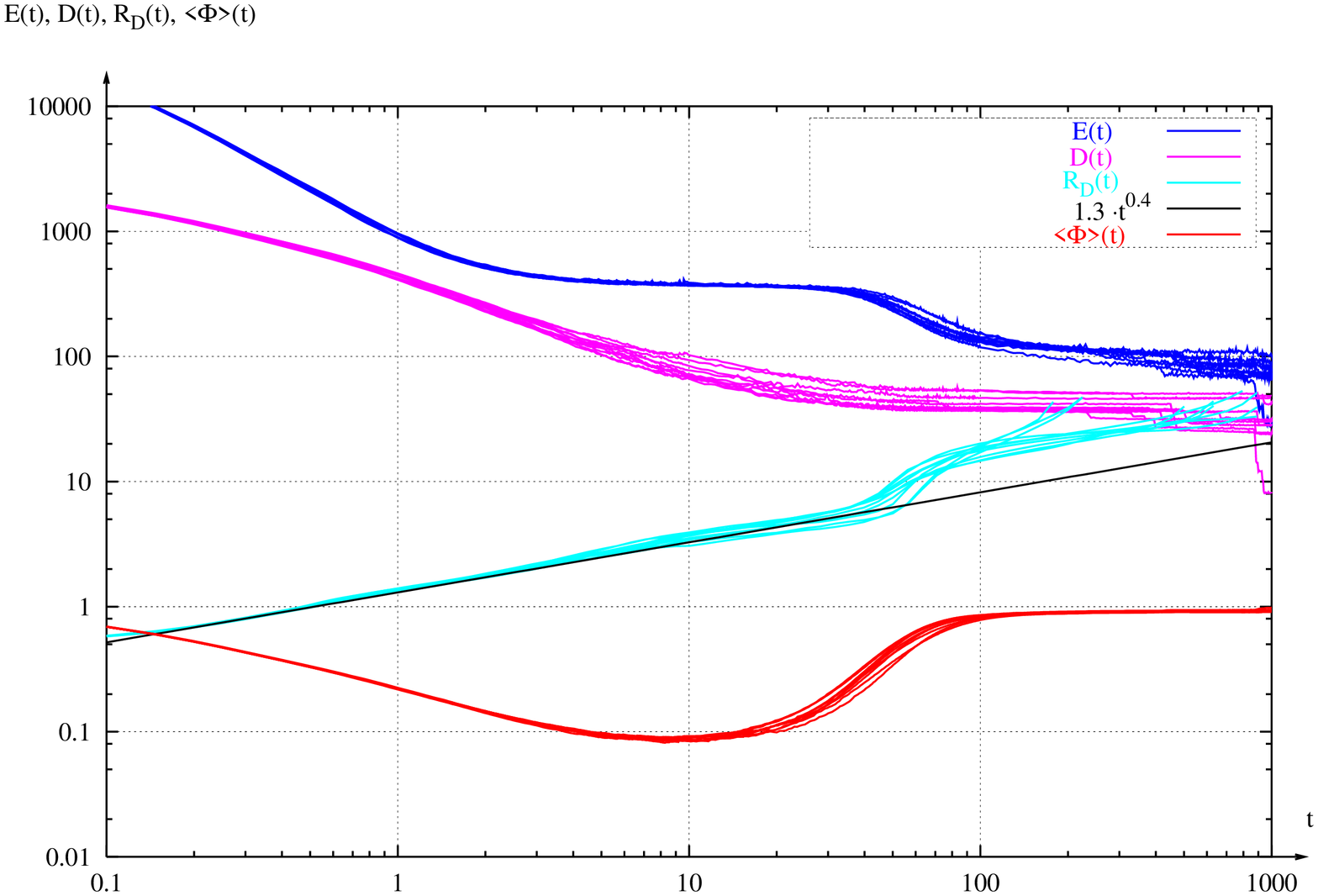}
\end{center}
\caption{ Total energy $E$, 'particle number' $D$, 
         'radius' $R_D$ of ordered domains, and the length $\langle \Phi
          \rangle$ of the field vector averaged over the whole lattice,  
for a few $B=0$-conserving evolutions (for $\lambda =1$, $\ell =
\sqrt{10}$, on a $120 \times 120$ lattice). The time is in units of the
relaxation time $\tau$. }
\label{serieB0}
\end{figure}

For a series of  $B$-conserving evolutions which start off from
different randomly chosen initial configurations, selected however for
winding number $B=0$, fig.\ref{serieB0} shows the time dependence of
the 'radius' $R_D$ of ordered domains, the total
energy $E$, the 'particle number' $D$ and the length $\langle \Phi
\rangle$ of the field vector averaged over the whole lattice.

\begin{figure}[h]
\begin{center}
\includegraphics[scale=0.34, angle=-90]{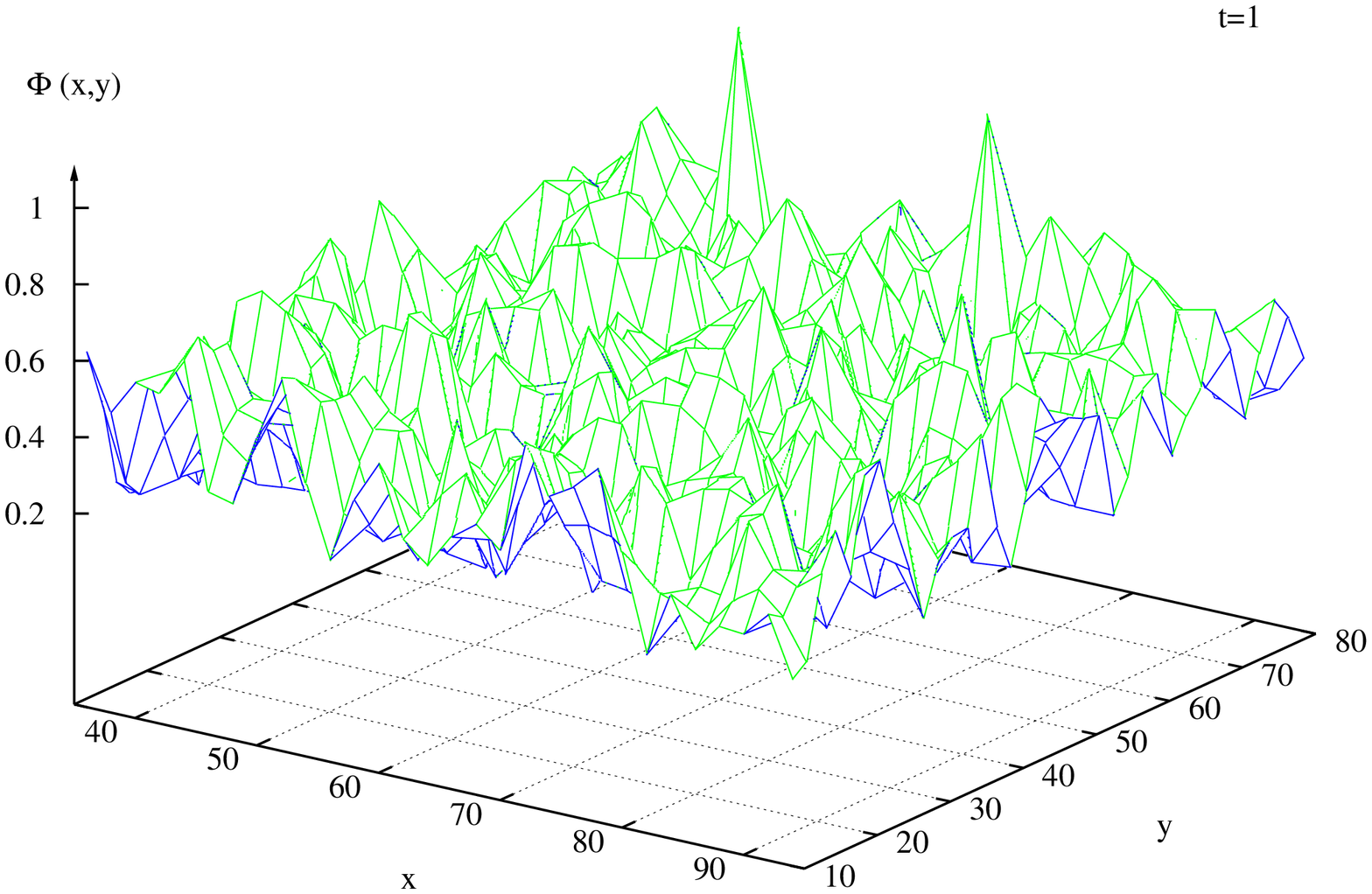}\includegraphics[scale=0.34, angle=-90]{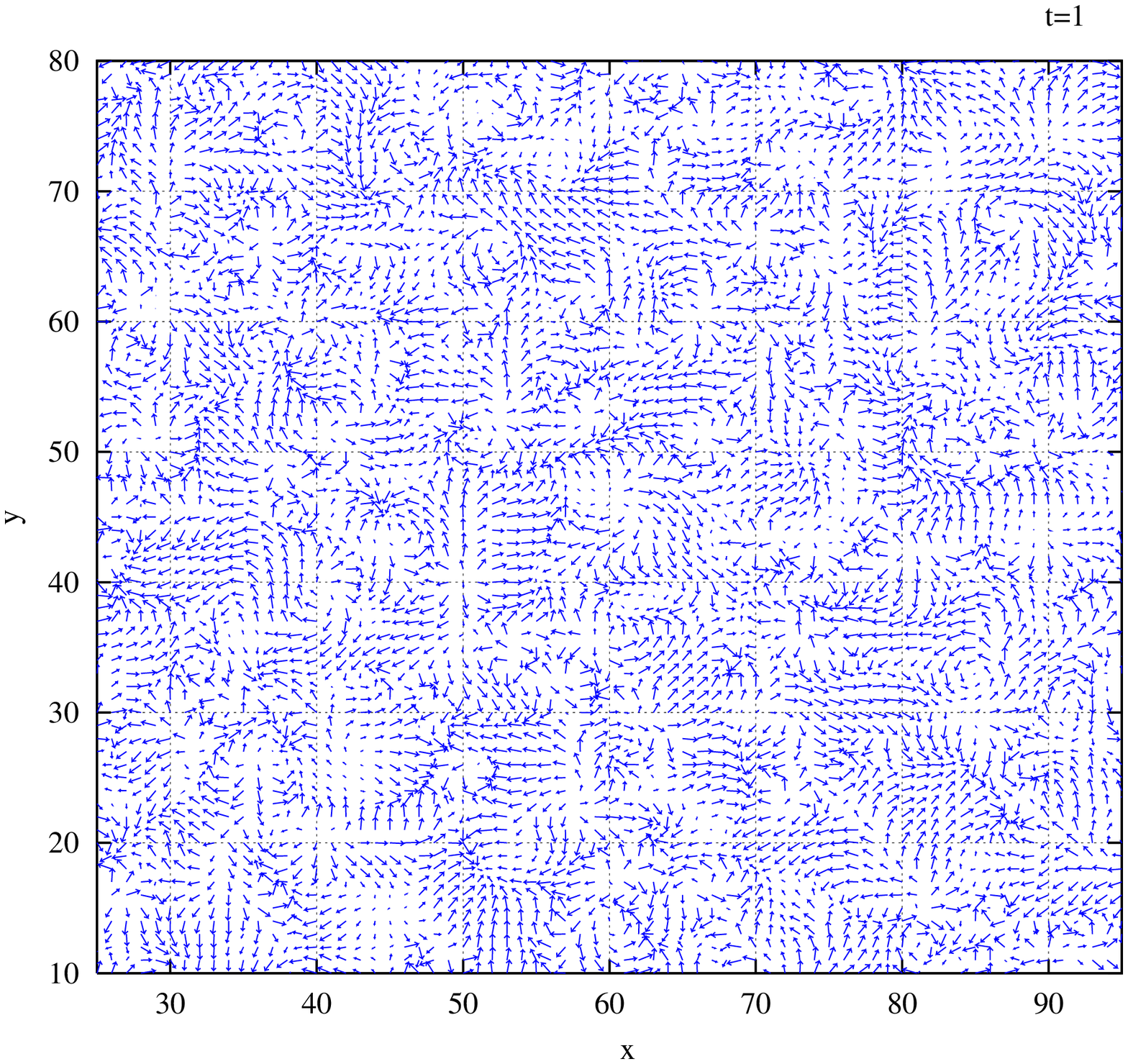}
\includegraphics[scale=0.34, angle=-90]{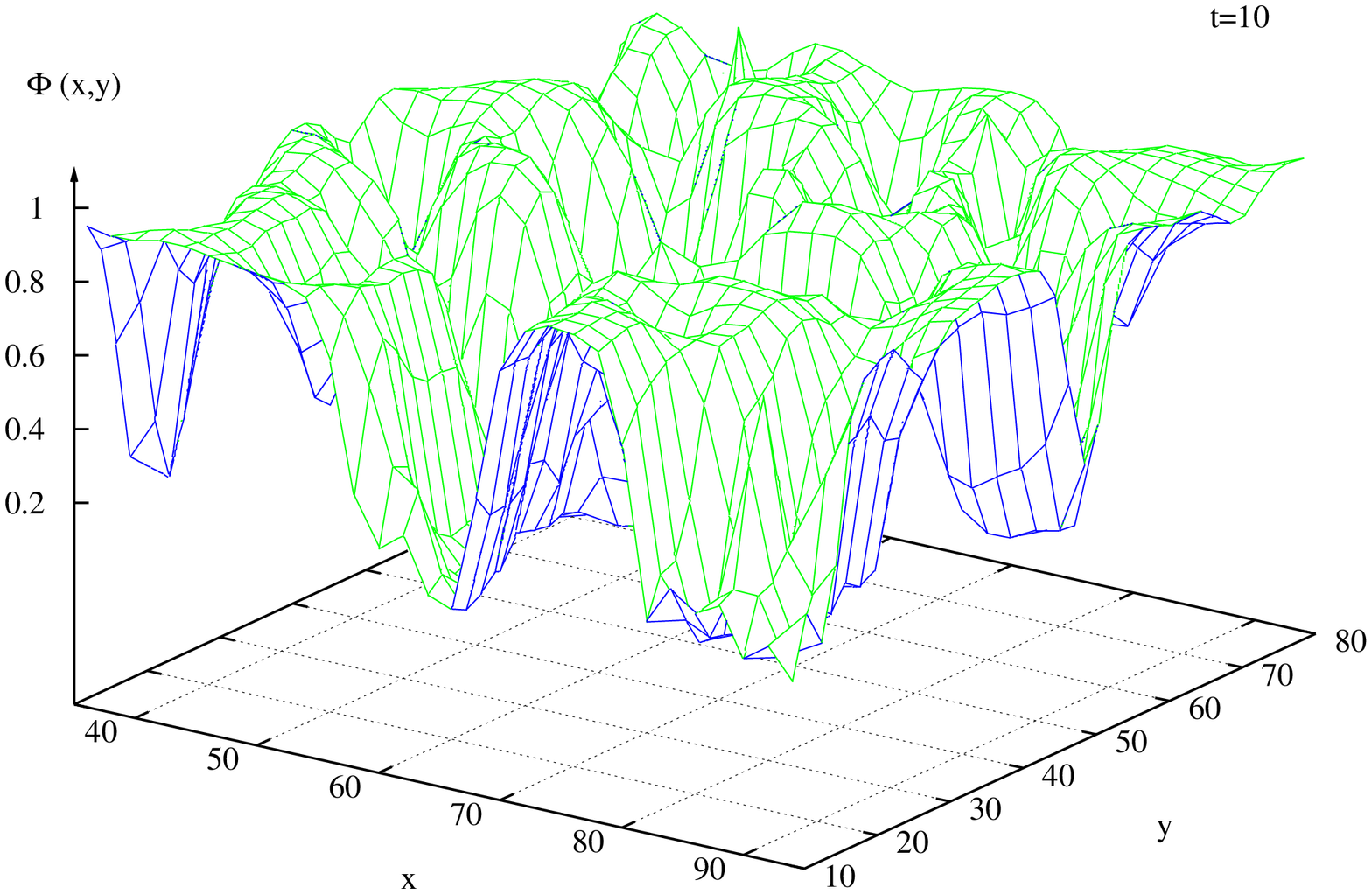}\includegraphics[scale=0.34,angle=-90]{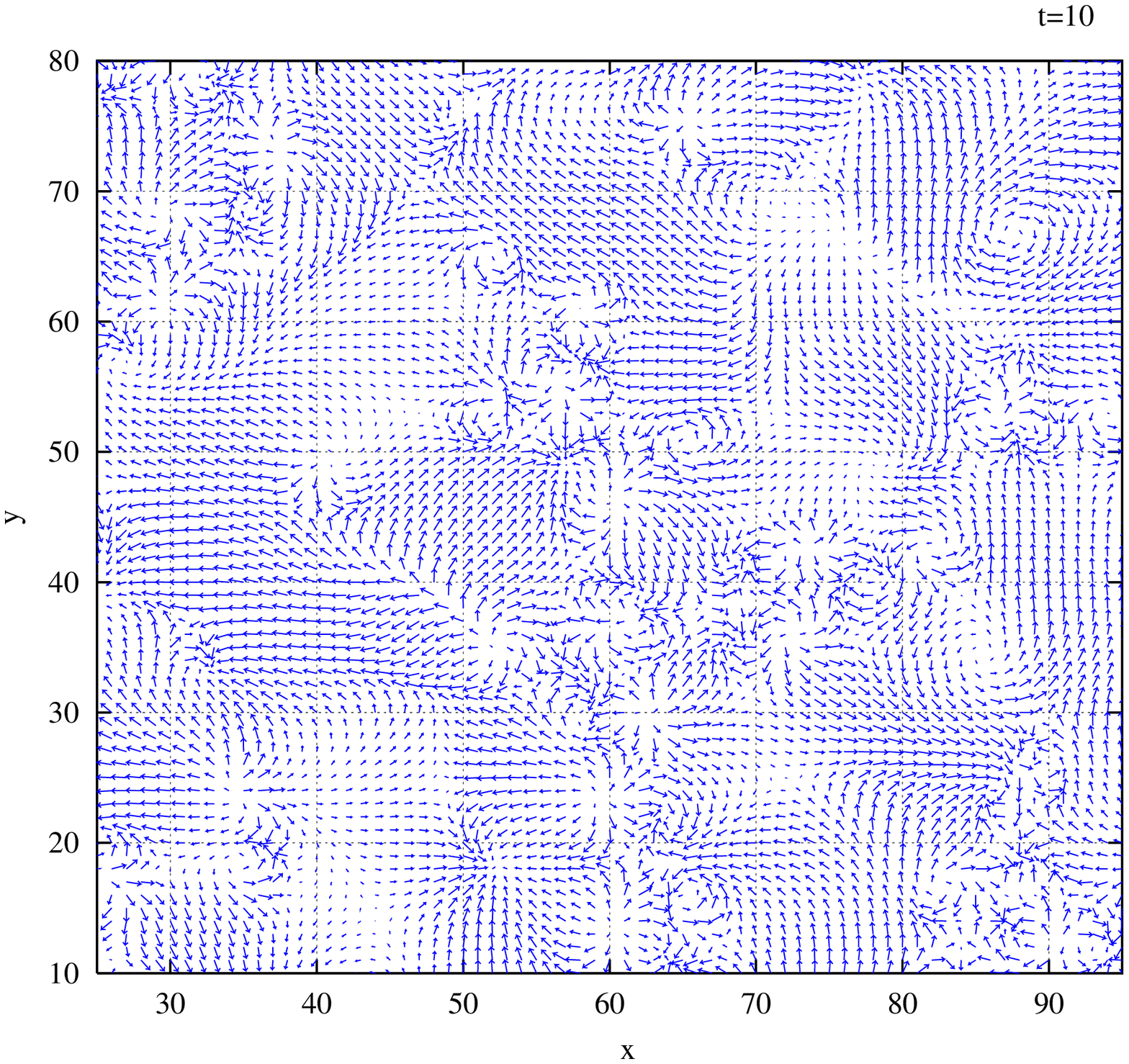}
\includegraphics[scale=0.34, angle=-90]{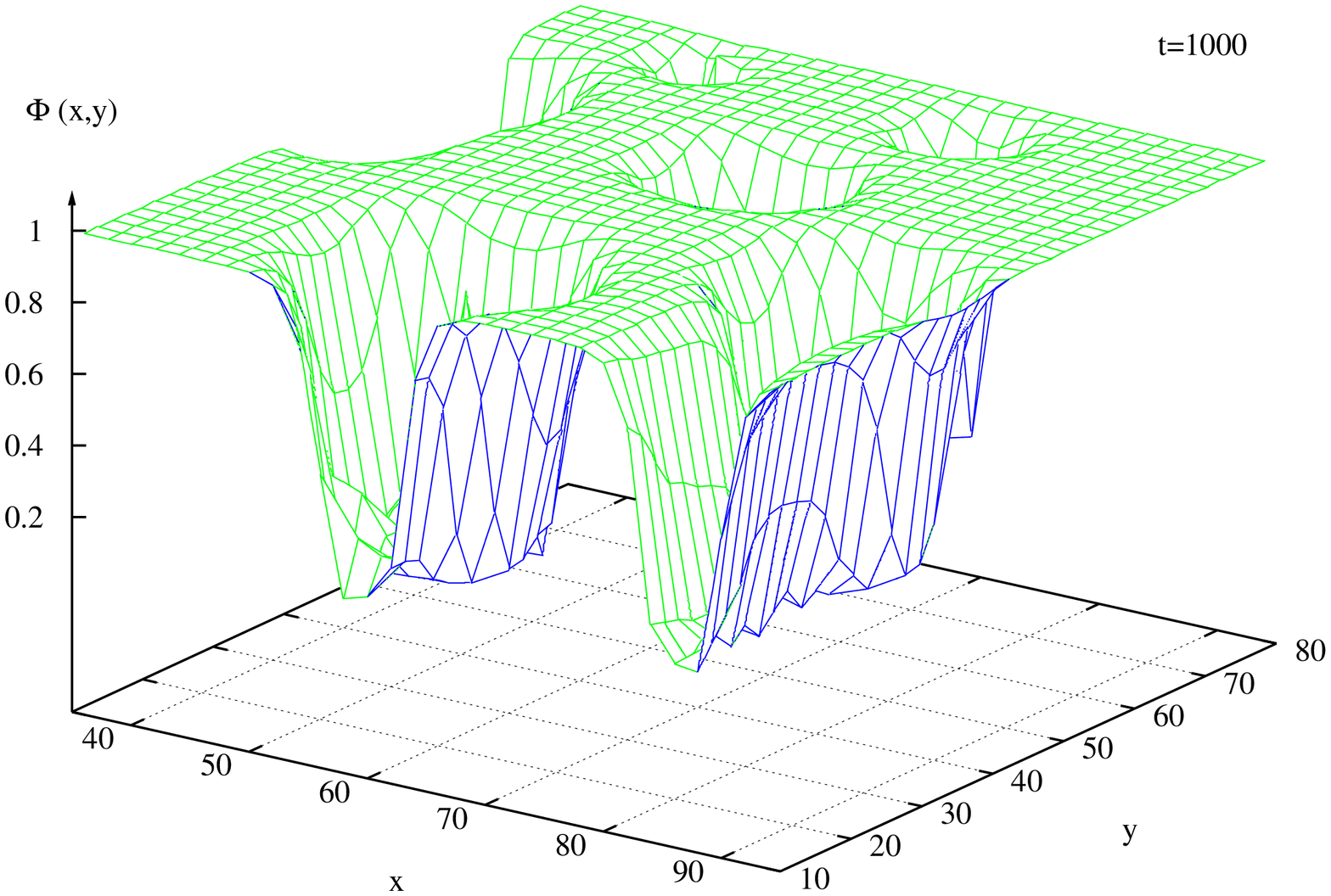}\includegraphics[scale=0.34, angle=-90]{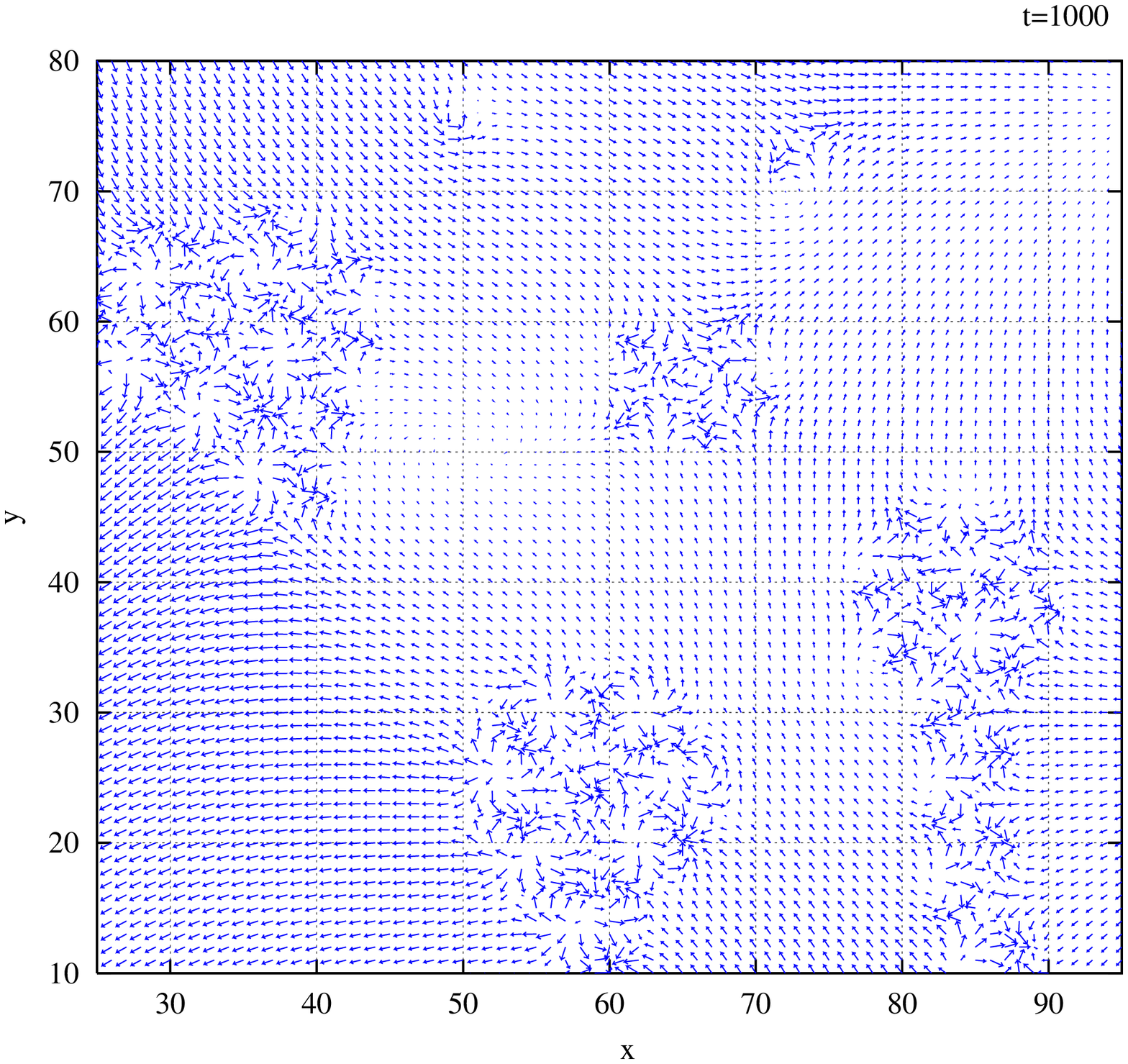}
\end{center}
\caption{Typical features of the field during different
stages of  an evolution  (for $\lambda =5$, with $B$ conserved at
$B$=0).
In the left column the lengths $\Phi$ are plotted over the spatial $x$-$y$ plane which provides a 3d-view of
the momentaneous bag structures. 
The right column presents the projection of the $\Bphi$-vectors at each lattice site
on the $\Phi_1-\Phi_2$ plane.}
\label{serieB0evolution}
\end{figure}
Figs.\ref{serieB0evolution} 
show the typical features of the field 
in an $70 \times 70$ section of a $150 \times 150$  lattice 
during different
stages of such an evolution
after $t=1$, $t=10$, and $t=1000$ relaxation time units.
\clearpage
\begin{figure}[t]
\begin{center}
\includegraphics[scale=0.45, angle=-90]{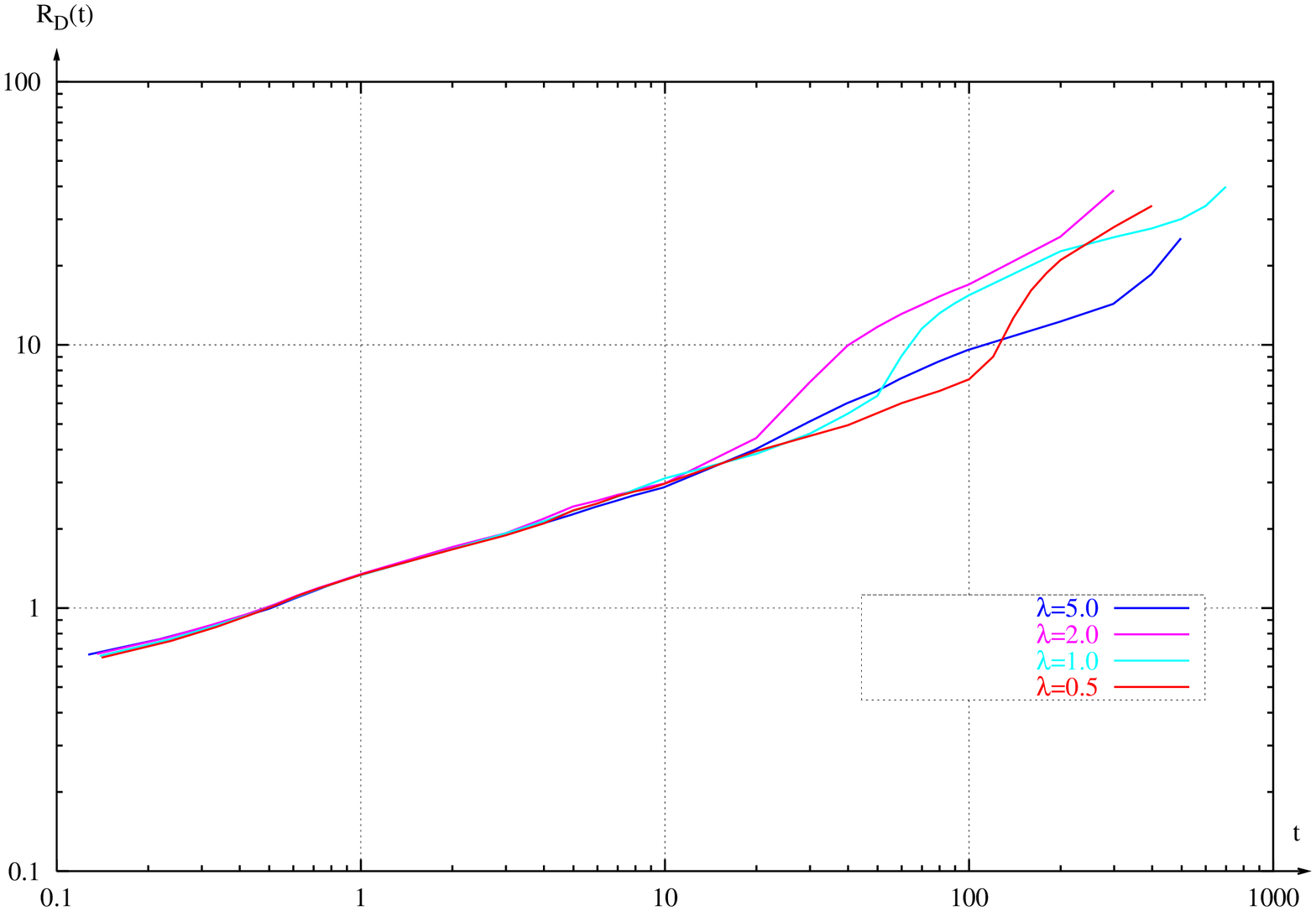}
\end{center}
\caption{Growth of the size $R_D$  for different
values of the coupling constant $\lambda$ ($\lambda = 0.5,1,2,5$)
for $B=0$-conserving evolutions on a $120 \times 120$ lattice.
}
\label{RD}
\end{figure}

Fig.\ref{RD} shows the growth of the size $R_D$  for different
values of the coupling constant $\lambda$ ($\lambda = 0.5,1,2,5$)
for $B=0$-conserving evolutions which start off from identical 
initial configurations.

One can distinguish three phases of the ordering process: 

i) During an initial 'relaxation' period which takes a few 
(relaxation-) time units, 
the lengths $\Phi$ of the field vectors initially rapidly
decrease and then vary around small values of about $0.1 f_0$,
the number of defects $D$ drops from its
starting value (which is of the order of $0.73 N^2/4$) by about one order
of magnitude, accompanied by a corresponding loss in total energy.
During this period the growth of $R_D$ closely follows a power
law 
\be
\label{power} 
R_D = a t^\alpha,  \mbox{    with  } \alpha \approx 0.4 \pm 0.01.
\ee
This exponent is
with good accuracy independent of the coupling constant $\lambda$, of
the initial configuration, and of the lattice size $N$
(as long as $N \gg 1$). It is also independent of the scale parameter
$\ell$. 
By the end of this period ordered domains extend over several lattice
units ($R_D\sim 5$). 

ii) The second phase ($10 < t < 100$, depending
on $\lambda$ and $\ell$) could be termed the 'roll-down' phase. It is
characterized by the increase of the (spatial-)average length $\langle
\Phi \rangle$ of the field vectors towards the vacuum value $f_0$. 
Actually, this roll-down process is rather slow; it takes of the order of
$\sim100$ relaxation time units for the space-averaged $\langle
\Phi \rangle$ to approach $f_0$.
Locally, this increase of $\Phi$ happens
only in the interior of ordered domains,
which results in the formation of numerous dense
and initially often connected bag structures located around the boundaries
of these domains. 
So, during this phase it is evidently the $\Phi^4$ potential which
drives the evolution. Therefore the onset of this
second phase depends on the scale parameter $\ell$ and on the coupling
strength $\lambda$. Large values of $\ell$ prevent the onset of this bag
forming process for a long time, so the angular alignment proceeds
further while $\langle \Phi \rangle$ is still small. This results in
much smaller 
total particle numbers when the bags finally are formed. Similarly,
this second phase starts earlier for larger values of $\lambda$.

Interestingly, the growth rate of the size of the ordered domains 
remains basically unaffected by this roll-down of $\Phi$: the increase
of the correlation functions proceeds monotonously through this
phase. There is, however, an effect on $R_D$ from the shoulder 
which appears in $C(R)$ for $5<R<10$ due to the
developing localized extended winding structures. If this shoulder 
passes through the halfmaximum which is used to define $R_D$ it leads
to a deviation from the power law
(\ref{power}) which is especially pronounced if the bag formation sets
in late, (i.e. for small values of $\lambda$ or large $\ell$), when the
size of the domains in which $\Phi$ approaches $f_0$ is larger.
When the bags are fully developed the ordered-domain size has increased
by about a factor of two, so the bags then are embedded in a patchwork
of ordered domains (with $\Phi \approx f_0$) which extend over 10-20
lattice units.

iii) The further development proceeds by the bags slowly moving around,
eating up smaller ones or uniting with others they meet on their way, or,
annihilating with others of opposite winding number. They assume 
sizes which correspond to the chosen scale $\ell$ and reflect the
partial winding number contained in their interior. So, naturally, the
evolution during this late period depends more on accidental features of 
the individual configurations as they have developed up to that point. 
However, it is interesting to note, that on the average, the growth of
the ordered domains (now with $\Phi=f_0$) again approximately follows the
power law (\ref{power}) with $\alpha$ around 0.4 (see fig.\ref{RD}).
 Finally, if $R_D$ has reached values near or
geater than $N/4$ the finite size of the lattice (with its periodic
boundary conditions) affects the long-range part of the correlation
functions resulting in a rapid artificial increase of $R_D$.

The optional filter on the total winding number $B$ allows to compare
evolutions with $B$ conserved at some initial value with others where
$B$ may jump freely during the course of the relaxation. 
As long as the actual values which $B$ takes on are
small as compared to $D$ there is almost no difference between
$B$-violating and $B$-conserving evolutions. For the major part of the
evolution the local particle-plus-antiparticle density is so high that
the evolution is dominated by annihilation processes, and occasional
unwinding jumps in $B$ by one or two units play no significant role. 
Only at late times ($t>500$) when $D$ has dropped below a few percent
of $N^2$ the possibility of spontaneous unwinding makes a noticeable
difference and it is accompanied by a correspondingly more rapid
increase of the size $R_D$ of ordered domains. 

\begin{figure}[t]
\begin{center}
\includegraphics[scale=0.33, angle=-90]{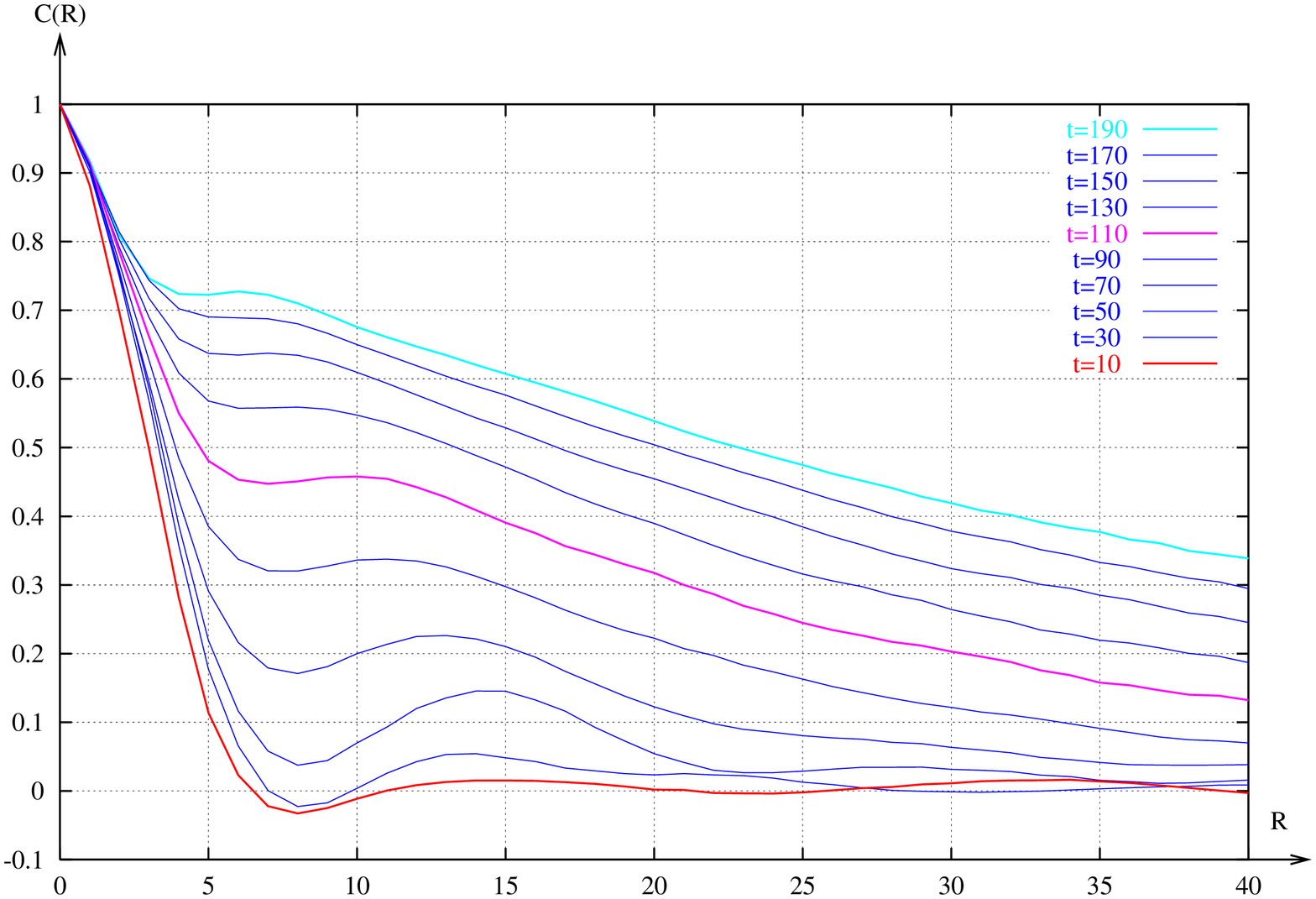}\includegraphics[scale=0.33, angle=-90]{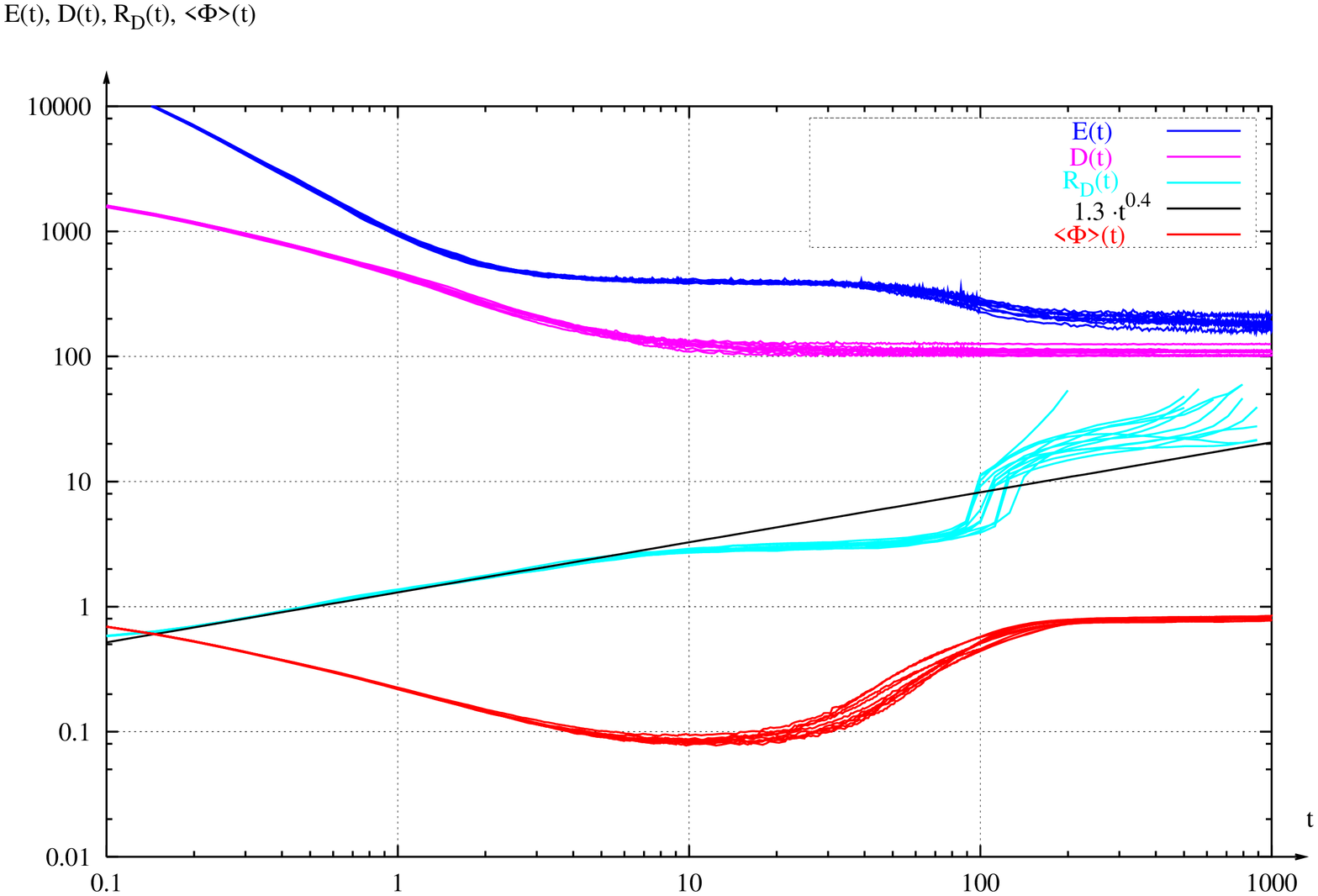}
\end{center}
\caption{Time evolution of the correlation function $C(R)$, and a series
of evolutions as in fig.\ref{serieB0}, however in a 'baryon-rich' 
environment where $B$ is selected and conserved at values $B>100$.}
\label{serieB100}
\end{figure}

\begin{figure}[h]
\begin{center}
\includegraphics[scale=0.34, angle=-90]{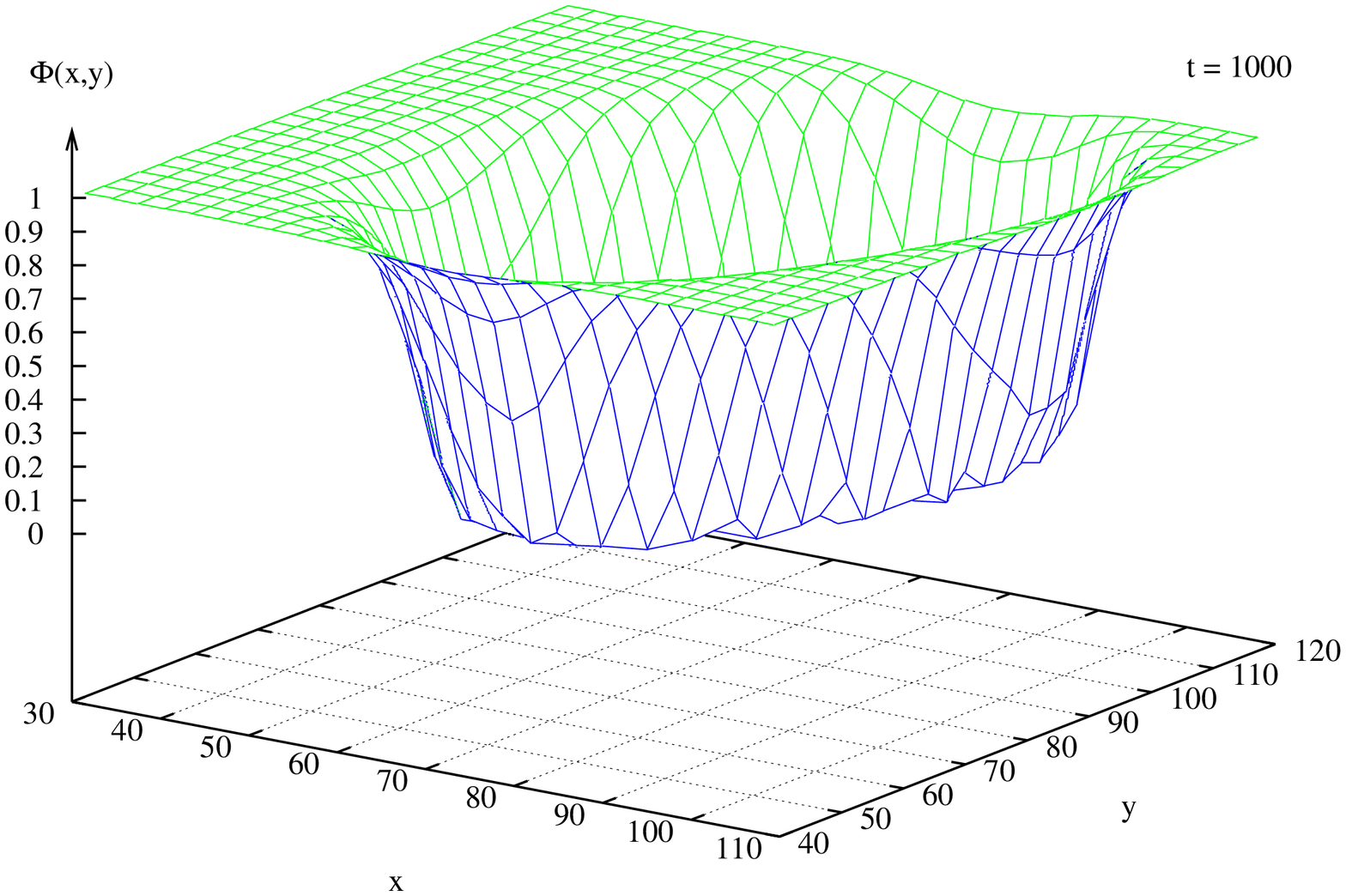}\includegraphics[scale=0.34, angle=-90]{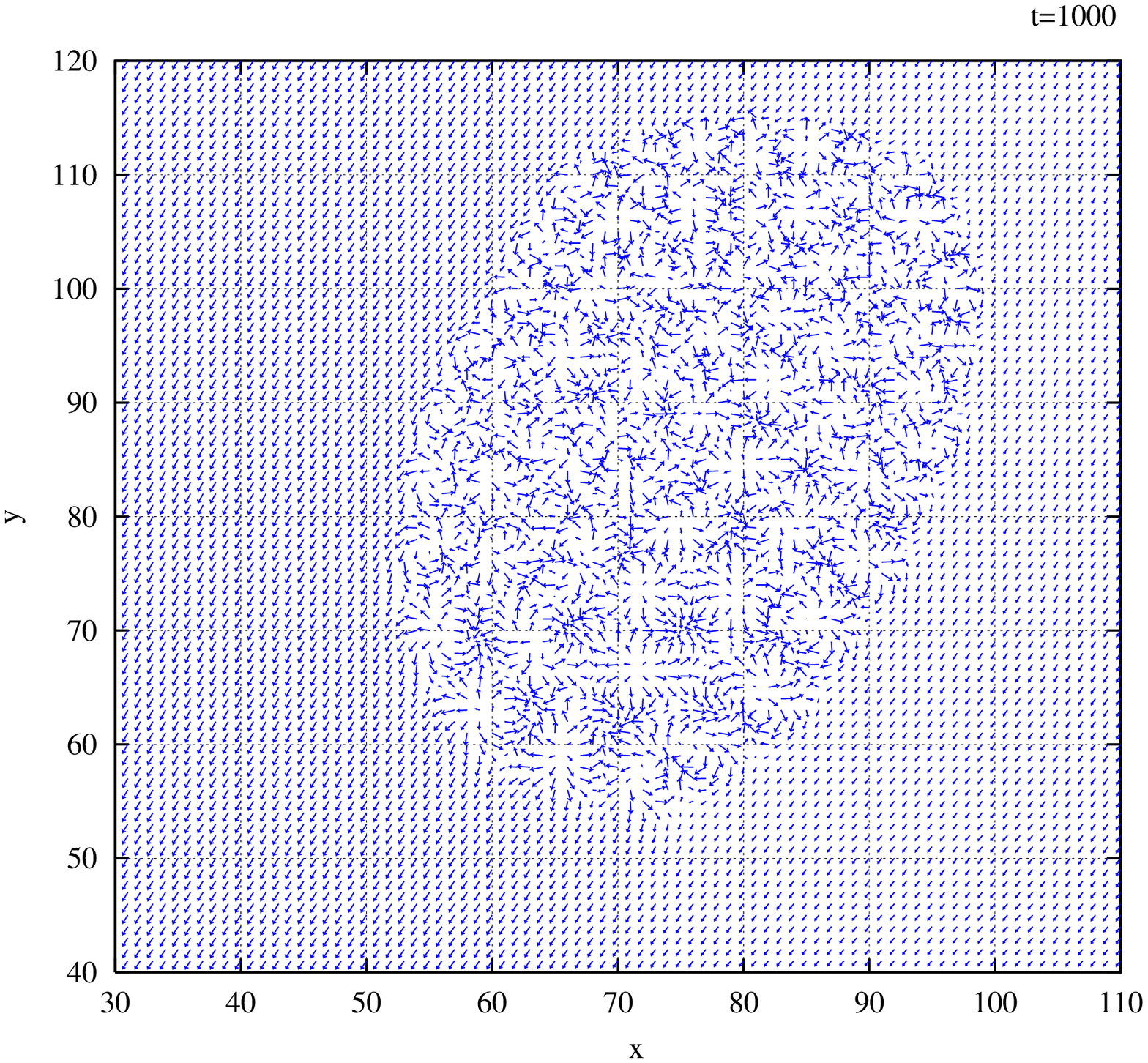}
\end{center}
\caption{
Typical example of the field in a baryon-rich environment (for $\lambda =1$,  with $B>100$)
 after $t=1000$ relaxation time units.
}
\label{serieB100evolution}
\end{figure}

On the other hand, by choosing a large initial value for $B$, the 
$B$-conserving relaxation
allows to study the formation of ordered domains in a 'baryon-rich'
environment. We present a series of such events in fig.\ref{serieB100} 
with $B>100$ on an $N=120$ lattice (for $\lambda=1$). In the early
stages as long as $D\gg B$ there is almost no difference as compared to
the $B=0$ case. However, for $t>10$ $D$ approaches the value of
$B$ fixed at $B>100$. At these times the value of $\Phi$ is still
much smaller than $f_0$ at most lattice points and the constraint on
winding number presents a severe obstacle for the growth of many aligned
domains, i.e. it causes a strong deviation from the power-law
(\ref{power}). In fact, the ordering proceeds in such a way that
only very few oriented domains start growing with $\Phi$
approaching $f_0$ in their interior. These few aligning domains squeeze
the regions  with nonvanishing winding (and very small $\Phi$)
into coherent large bags, such that the whole space becomes
separated into large aligned areas and large bags. This structure is
reflected in the correlation functions as a very pronounced shoulder
in the range around $5<R<10$ which rises with increasing time due to
the further increase of long range correlations (cf. fig.\ref{serieB100}).
Evidently, near $t \sim 100$ this leads to
an almost instantaneous strong increase in $R_D$, which thus appears 
more as a consequence of the definition of $R_D$ as the half-maximum
distance rather than an actual abrupt increase of the size of aligned
areas. The further
development then is characterized by the formation of one large bag which
comprises almost all of the winding number (cf. fig.\ref{serieB100evolution}),
 while alignment in the 
surrounding 'vacuum' in most cases progresses slowly according to
(\ref{power}) with $\alpha\sim 0.4$ before finite-lattice-size effects
set in for $t \sim 500$. 

Apart from fixing the spatial extent of the finally formed bags 
the scale parameter $\ell$ affects during the early stages of the
evolution the duration of the first period where the bags are not yet fully
developed. Thus it allows to monitor the total particle number present
at the time of bag formation. This is an interesting aspect for
evolutions where one does not consider a sudden quench but allows $f_0$ to
change with the temperature of the system. 
According to eq.(\ref{rescale}) this transforms into changing $\ell$
with time.

\section{Conclusion}

We have presented here numerical simulations of phase ordering 
for a 3-vector field through spontaneous symmetry breaking in two
spatial dimensions, with specific attention to the interplay
between the aligning process and the formation of stable extended
topological structures. For that purpose
the effective action is chosen in such a way that the localized
topological defects which necessarily accompany the formation of
randomly oriented aligned domains are stabilized with a definite size
as ordered localized structures embedded in the aligning field.
The stabilizing terms do not break the $O(3)$ symmetry explicitly.
For definiteness we have only considered the sudden quench scenario
and assumed overdamped dynamics.
Apart from the interesting features of the coarsening transition 
such processes may serve as models for the spontaneous
creation of extended particles and antiparticles or clusters of those
out of a hot random field ensemble. 

Three aspects are of peculiar interest:
The growth rate of the size of aligned domains follows a power law
with an exponent of approximately 0.4, which persists through the early
relaxation phase and the subsequent roll-down phase and is apparently
quite independent of the stabilizing terms in the action. This exponent
is in agreement with previous results found in models without (or with
only one of the) stabilizing terms for the average defect-defect
separation~\cite{Rut} and the spin-spin correlation~\cite{Zapo}. 
The formation
of stable defects is most prominently reflected in the shape of the
equal-time angular correlation functions. During these earlier phases
the ordering process is dominated by defect-antidefect annihilation 
which reduce the initial particle number by up to two orders of
magnitude before the remaining defects slowly take on their stable
conformation. By that time (which is of the order of several hundred
relaxation times) the final alignment of the remaining few large
disoriented domains depends sensitively on the accidental spatial
configuration of the few surviving extended particle clusters.

The scale parameter $\ell$ which determines the size of the resulting
defects does not affect the early relaxation period but it has a
pronounced influence on the subsequent evolution:
larger values of $\ell$ suppress the increase of $\Phi$ 
towards its vacuum value $f_0$ over larger spatial areas and thus
delay the onset of the roll-down phase.
This leads to a reduction of the final density of particles plus
antiparticles when the bags finally emerge. 
Through (\ref{rescale}) $\ell$ is directly related to $f_0$ which
implies sensitivity of the final total particle number to the quench
velocity. 

The third interesting feature concerns $B$-conserving evolutions in
an environment with large values of $B$. Similar to the case of large
$\ell$ the onset of the roll-down is delayed until most of the
possible annihilations have taken place, i.e. until the total
particle number approaches the fixed winding number. 
The subsequent growth of only a few aligned domains squeezes regions
with nonvanishing winding density into large coherent particle clusters
which fill the interior of large bags. 

Extending the present considerations to the 3D-$O$(4) model appears
as a challenging task in view of the ongoing discussion of the chiral
phase transition, the formation of disoriented chiral domains, and
baryon-antibaryon production in the cooling of hot hadronic plasma.
It may help to establish further links between assumed quench
scenarios and signatures to be expected from emitted particles.


\begin{thebibliography}{99}
\itemsep=0.5cm

\bibitem{Burgess} C. P. Burgess,  {\em Phys.Rept.} {\bf 330}, 193 (2000).

\bibitem{ChPT} S. Weinberg, {\em Physica} {\bf 96A}, 327 (1979);\\
J. Gasser and H. Leutwyler, {\em Ann. Phys. (N.Y.)}{\bf 158}, 142 (1984). 

\bibitem{Anselm} A.A. Anselm, {\em Phys. Lett.} {\bf B217}, 169 (1988); 
A.A. Anselm and M.G. Ryskin, {\em Phys. Lett.} {\bf B266}, 482 (1991);\\
J.P. Blaizot and A. Krzywicki, {\em Phys. Rev.} 
{\bf D46}, 246 (1992); {\em Phys. Rev.} {\bf D50}, 442 (1994);\\
J.D. Bjorken, {\em Int. J. Mod. Phys.} {\bf  A7}, 4189
(1992); {\em Acta Physica Polonica} {\bf B23}, 561 (1992);
{\em Disoriented chiral condensate}, 
Proc. Workshop on Continuous advances in QCD, Minneapolis
 1994, and SLAC-PUB-6488 (1994);\\
S.Gavin, {\em Nucl. Phys.} {\bf A590}, 163c (1995).

\bibitem{Ellis} T.A. DeGrand, {\em Phys. Rev.} {\bf D30}, 2001 (1984); \\
J. Ellis and H. Kowalski, {\em Phys. Lett.} {\bf B214}, 
161 (1988); J. Ellis, U. Heinz, and H. Kowalski, {\em Phys. Lett.} 
{\bf B233}, 223 (1989); J. Ellis, M. Karliner, and H. Kowalski, 
{\em Phys. Lett.} {\bf B235}, 341 (1990).

\bibitem{QHE}
S.L. Sondhi, A. Karlhede, S.A. Kivelson, and E.H.Rezayi, Phys. Rev. {\bf
B47}, 16419 (1993). 

\bibitem{SkyWi} T.H.R. Skyrme,  Proc. R. Soc. {\bf A260}, 127 (1961);\\
E. Witten,  Nucl. Phys. {\bf B223}, 422,433 (1983).

\bibitem{Zapo} M. Zapotocky and W.J. Zakrzewski,  Phys. Rev. {\bf
E51}, R5189  (1995);\\
A.D. Rutenberg, W.J. Zakrzewski, and M. Zapotocky,
Europhys. Lett.  {\bf 39}, 49 (1997).

\bibitem{Rut} A.D. Rutenberg, Phys. Rev. {\bf E51}, R2715 (1995);\\
G. J. Stephens, Phys. Rev. {\bf D61}, 085002 (2000).    

\bibitem{Ho} G. Holzwarth,  Phys. Rev. {\bf D59}, 105022 (1999).  

\bibitem{Hans97} H. Walliser,  Phys. Rev. {\bf D56}, 3866 (1997).

\bibitem{BP} 
A.A. Belavin and A.M. Polyakov,  JETP Lett. {\bf 22}, 245 (1975).

\bibitem{Bjorken} J.D. Bjorken, Phys. Rev. {\bf D27}, 140 (1983).

\bibitem{Kibble} T.W.B. Kibble, {\em J.Phys.}{\bf A9}, 1387 (1976);\\
N.H. Christ, R. Friedberg and T.D. Lee, 
{\em Nucl.Phys.} {\bf B202}, 89 (1982).

\end{thebibliography}
\end{document}